\documentclass[manuscript]{aastex}
\usepackage{natbib}
\usepackage[usenames,dvipsnames,graphics,graphicx]{xcolor}
\usepackage{lscape}
\begin{document}


\title{ALMA High Angular Resolution Polarization Study; An Extremely Young Class 0 Source, OMC-3/MMS 6}

\author{Satoko Takahashi$^{1, 2, 3, 4}$, Masahiro N. Machida$^{5}$, Kohji Tomisaka$^{3,6}$, 
Paul T. P. Ho$^{4,7}$, Edward B. Fomalont$^{1,8}$, Kouichiro Nakanishi$^{3,6}$, and Josep Miquel Girart$^{9,10}$}
\affil{ 
$^{1}$Joint ALMA Observatory, Alonso de C{\'{o}}rdova 3107, Vitacura, Santiago, Chile; satoko.takahashi@nao.ac.jp, \\ 
$^{2}$NAOJ Chile Observatory, Alonso de C{\'{o}}rdova 3788, Oficina 61B, Vitacura, Santiago, Chile, \\
$^{3}$Department of Astronomical Science, School of Physical Sciences, SOKENDAI (The Graduate University for Advanced Studies), Mitaka, Tokyo 181-8588, Japan\\
$^{4}$Academia Sinica Institute of Astronomy and Astrophysics, P.O. Box 23-141, Taipei 106, Taiwan, \\
$^{5}$Department of Earth and Planetary Sciences, Faculty of Sciences, Kyushu University, Fukuoka 812-8581, Japan, \\
$^{6}$National Astronomical Observatory of Japan, Mitaka, Tokyo 181-8588, Japan, \\
$^{7}$East Asian Observatory, Hilo 96720, HI, USA, \\
$^{8}$National Radio Astronomy Observatory, Charlottesville, VA 22903, USA,  \\
$^{9}$Institut de Ci$\rm{\grave{e}}$ncies de l’Espai (ICE, CSIC), Can Magrans s/n, E-08193 Cerdanyola del Vall$\rm{\grave{e}}$s, Catalonia, Spain  \\
$^{10}$Institut d’Estudis Espacials de de Catalunya (IEEC), E-08034 Barcelona, Catalonia, Spain}

   \date{December 5, 2018}

\begin{abstract}

Using the $\approx$15km ALMA long baselines, 
we imaged the Stokes $I$ emission and linearly polarized intensity ($PI$) 
in the 1.1-mm continuum band of a very young 
intermediate-mass protostellar source, MMS 6, 
in the Orion Molecular Cloud-3. 
The achieved angular resolution, 
$0''.02{\times}0''.03$ ($\approx$10 AU), 
shows for the first time 
a wealth of data on the dust emission polarization in the central 200 AU of a protostar. 
The $PI$ peak is offset to the south-west (SW) by $\approx$20 AU with 
respect to the Stokes $I$ peak. 
Its polarization degree is 11\% 
with its $E$-vector orientation of P.A.${\approx}135^{\circ}$. 
A partial ring-like structure with a radius of $\approx$80 AU 
is detected in $PI$ but not in the Stokes $I$. 
NW (north-west) and SE (south-east) parts of 
the ring are bright with a high polarization degree of $\gtrsim$10\%, 
and their $E$-vector orientations are roughly orthogonal to those 
observed near the center. We also detected arm-like polarized structures, 
extending to 1000 AU scale to the north, with the $E$-vectors 
aligned along the minor axis of the structures. 
We explored possible origins of the polarized emission comparing with magnetohydrodynamical 
(MHD) simulations of the toroidal wrapping of the magnetic field. 
The simulations are consistent with the $PI$ emission in the ring-like 
and the extended arm-like structures observed with ALMA. 
However, the current simulations do not completely reproduce 
observed polarization characteristics in the central 50 AU. 
Although the self-scattering model can explain 
the polarization pattern and positional offset between 
the Stokes $I$ and $PI$, this model is not able to reproduce 
the observed high degree of polarization.

\end{abstract}

\keywords{polarization -- starts:individual(OMC-3/ MMS 6) -- stars: formation -- ISM: jets and outflows}

\section{Introduction}

Magnetic fields are one of the key elements that regulate star
formation \citep{1987ARA&A..25...23S,2012ARA&A..50...29C}, and 
organized magnetic fields are often observed in parsec-scale dense
molecular clouds and cores. 
Even when the Lorentz force is insufficient to balance the gravity 
and to prevent contraction of the core \citep{2012ARA&A..50...29C}, 
the magnetic fields are still important in the star formation process. 
For instance, in a large scales, the magnetic field 
seems to produce disk-like structure, 
which is called ``pseudo disk''  (\citealt{1987ARA&A..25...23S,1988PASJ...40..593N,1993ApJ...417..243G}). 
In the vicinity of the protostar, the Lorentz force plays 
a critical role for 
launching outflows and jets
\citep{2002ApJ...575..306T,2006ApJ...641..949B,2008ApJ...676.1088M,
2010A&A...510L...3C,2010ApJ...714L..58T},
which are commonly observed in the star forming regions (e.g.,
\citealt{2007prpl.conf..245A} and references therein). A strong
magnetic field removes the angular momentum from the disk via magnetic
braking (e.g., \citealt{2008ApJ...681.1356M}) and produce outflows (e.g.,      
\citealt{2008ApJ...676.1088M,2011PASJ...63..147T}), while the Ohmic
dissipation promotes the disk formation and            
growth (e.g., \citealt{2011PASJ...63..555M}). These effects determine
the properties of a rotationally supported disk in the early stage of 
the protostellar evolution.

A method to study the magnetic field structures in the protostellar core is to
observe the linearly polarized thermal emission from magnetically aligned dust grains (e.g.,
\citealt{1949Sci...109..461S,1951ApJ...114..206D,1988QJRAS..29..327H,2007JQSRT.106..225L,1998ApJ...502L..75R,2009ApJ...707..921R,2014ApJ...780L...6R,2003ApJ...598..392L,2014ApJS..213...13H,2017ApJ...847...92H,2006Sci...313..812G,2009Sci...324.1408G,2013ApJ...772...69G,2013ApJ...763..135T,2014ApJ...792..116Z,2016ApJ...825L..15C,2018ApJ...855...39K,2018MNRAS.tmp..552M,2018ApJ...856L..27G,2018A&A...616A..56A,2018ApJ...854...56L} and references therein). 
Since the degree of linear polarization observed 
in the star forming cores is typically $\lesssim$5\%, 
only a sensitive array such as ALMA can image dust polarization 
for faint sources and in their most internal parts ($\lesssim$ a few $\times$100 AU). 
With the ALMA sensitivity and its angular resolution 
that enables to probe nearby protostars 
at 10 AU scales, magnetic field structures 
can be traced in detail and measured magnetic 
field structures can be directly 
compared with magnetohydrodynamical (MHD) simulations 
(e.g., \citealt{2008ApJ...676.1088M,2011PASJ...63..147T,2014ApJ...793..130L}). 
Such observations are key to study the formation processes 
of rotationally supported disks and the launching mechanism of the jet and outflow.

At small scales ($\lesssim$ a few$\times$ 100 AU), 
dust polarization is not always associated with the magnetic field 
and could also be related to other mechanisms such as 
the self-scattering \citep{2015ApJ...809...78K,2016MNRAS.460.4109Y} 
and dust grain alignment due to the anisotropic radiation 
\citep{2017ApJ...839...56T}. 
Each mechanism produces different orientations 
of the polarization vectors and different dependence of 
the degree of linear polarization in frequency.

We observed the brightest Class 0 intermediate-mass (IM) protostellar core, MMS 6, 
located in the Orion Molecular Cloud-3 region (OMC-3; $d$=414 pc by \citealt{2007A&A...474..515M} or $d=$388 pc by \citealt{2017ApJ...834..142K})
\footnote{In this paper, the distance to the object, $d$=414 pc, is adopted.}. 
MMS 6 has a bolometric luminosity of $L_{\rm{bol}}<$ 60 $L_{\odot}$ and 
a core mass of $M_{\rm{core}}=$30 $M_{\odot}$ 
\citep{1997ApJ...474L.135C}. \citet{2012ApJ...752...10T} 
detected a massive gas envelope (0.29 $M_{\odot}$), 
the presence of hot gas ($\gtrsim$52 K), and
extremely high column density
(\hbox{$N_{\rm{H_2}}=$2.1$\times$10$^{25}$ cm$^{-2}$}), in the central
120 AU. In addition, \citet{2012ApJ...745L..10T} detected an extremely
compact ($\approx$1000 AU) and collimated molecular outflow associated
with MMS 6.  These results imply that MMS 6 is one of the youngest IM core. 
MMS 6 is a bright source and relatively close, so it is a unique target to
observe the polarized emission at the ALMA highest angular resolutions.

We here report on new ALMA high angular resolution observations of the polarization of the dust continuum emission in the protostellar core MMS 6 at sub-arc ($0''.02{\times}0''.03$) resolution corresponding to 10 AU for the adopted distance of MMS 6. 
Previous studies indicate that MMS 6 is a clear case of a magnetically 
dominated region. The large-scale magnetic field orientations in the 
northern part of OMC-3, where MMS 6 is located, show an organized
magnetic field structure that is smoothly connected between 
the size scale of filamentary cloud, core, and envelope \citep{2000ApJ...531..868M,2001ApJ...562..400M,2005ApJ...626..959M,
2010ApJ...716..893P,2014ApJS..213...13H}. 

In different star forming regions, 
ALMA polarization observations toward 
protoplanetary disks have been extensively performed 
enabling detailed discussions of the origin of their emission 
(e.g., \citealt{2016ApJ...831L..12K,2018ApJ...860...82H,
2018ApJ...865L..12B,2018ApJ...864...81O,2019MNRAS.482L..29D}). 
For the earlier evolutionary stage (i.e., Class I/0), 
many observational results are also available today (\citealt{2017ApJ...851...55S,2017ApJ...844L...5K,
2017ApJ...847...92H,2018MNRAS.tmp..552M,2018ApJ...855...92C,
2018A&A...616A..56A,2018arXiv180507348K,
2018ApJ...854...56L,2018ApJ...859..165S,2018ApJ...861...91H,2018ApJ...856L..27G}), 
showing a high degree in variations in the origin of the polarized emission and  
displaying complex morphologies. 

The new ALMA data allow us to study the polarization and 
magnetic field structures at the center of a protostellar core, 
MMS 6/OMC-3 at the highest angular resolution as possible with current 
ALMA at the observed frequency, and to better understand 
the origin(s) of the dust polarized 
emission, characterize its nature, 
and compare the results with other Class 0/I sources. 

The paper is organized as follows. 
The dust polarization observations and 
the molecular outflow observations are described in Section 2. 
The total intensity and the polarized intensity distributions, 
from 10 AU to a few $\times$1000 AU scales, are shown 
and described in Section 3. 
Origins of the ALMA dust polarized emission in MMS 6 
and comparisons with the MHD simulation 
are discussed in Section 4. 
Finally, concluding remarks as well as future prospects are given in Section 5.

\section{OBSERVATIONS AND DATA REDUCTION}

\subsection{The 1.1 mm Continuum Polarimetric Observations}

The ALMA observations of MMS 6 were obtained through 
the science project 2015.1.00341.S (P.I. S. Takahashi) 
using two different configurations. 
The high angular resolution observations were made 
in Cycle 3 on 2015 October 29 with the 16-km ALMA configuration 
in two consecutive observing blocks, 
and the lower angular resolution observations 
were made in Cycle 4 on 2016 October 9 and 11 
with the 3.6-km ALMA configuration in five observing blocks. 
The phase center of all of the observations were 
R.A. (J2000) = 5$^h$35$^m$23$^s$.4200, 
decl. (J2000) =−05$^{\circ}$01$'$30$''$.350. 
Observing parameters associated 
with the three observations are listed in Table 1. 
Both high and low angular resolution observations 
had about 40 minutes on-source time with about forty 12-m 
diameter antennas. The low and high angular resolution data sets 
cover projected baselines between 16 k$\lambda$ 
and 3200 k$\lambda$ and between 76 k$\lambda$ 
and 14700 k$\lambda$, respectively. 
The two data sets are respectively insensitive 
to structures more extended than $1''.0$ and $0''.22$ 
at the 10\% level \citep{1994ApJ...427..898W}. 
Four spectral windows (SPWs) with 1.875 GHz bandwidth, 
centered at 256, 258, 272, and 274 GHz 
are allocated with the time division mode  
for this polarization experiment, 
giving a total continuum bandwidth of 7.5 GHz. 

The Common Astronomy Software Application (CASA; 
\citealt{2007ASPC..376..127M}) version 4.5.0 was used 
for the standard ALMA data reduction. 
The calibration scripts were provided by the observatory. 
The calibration steps include 
(1) correcting the gains associated 
with the variable receiver and sky noise and phases 
associated with the water vapor along the line of site, 
(2) providing the flux density scale with an observation of calibrator 
of known flux density, (3) removing the amplitude and 
phase frequency dependence (bandpass) for each SPW, 
(4) removing the amplitude and phase temporal 
dependence using a phase calibrator 
within a few degrees of MMS 6, 
(5) removing instrumental polarization using a bright polarized 
calibrator, J0522-3627, which was observed every 25 minutes 
during each execution, 
and (6) making appropriate data flagging as calibrations progress. 
The calibrated data from the two 16-km high resolution 
experiments were combined into the high resolution data set. 
The data from the five 3.6-km low resolution experiments were 
combined into the low resolution data set. 
The data points in the combinations were weighted by 
their theoretical SNR which gives best theoretical SNR 
in the combined data sets.

Then CLEANed images were made using the CASA task ``clean''. 
The Briggs weighting with robust parameter of 0.5, 
and natural weighting were used for the high 
and low angular resolution final images, respectively. 
In order to reduce residual phase errors and
improve the dynamic range of the images, self-calibrations
have been applied for the low angular resolution image. 
This process improved the maximum dynamic range by a factor of 1.8. 
The resulting synthesized beam sizes are 
0$''$.03 $\times$ 0$''$.02 (P.A.= 43$^{\circ}$) 
and 0$''$.15 $\times$ 0$''$.14 (P.A.= $-80^{\circ}$) for 
the high angular resolution and low angular resolution images, respectively. 
The achieved rms noise level (1 $\sigma$) for the high angular 
resolution and low angular resolution images 
(for Stokes $I$, $Q$, and $U$) are 63, 21, 21$\mu$Jy beam$^{-1}$ 
and 130, 20, 20 $\mu$Jy beam$^{-1}$, respectively. 
The Stokes $Q$ and $U$ image rms is near the expected thermal 
noise level. However, the Stokes $I$ image 
is limited by small residual phase errors that could 
not be calibrated using phase referencing. 
Still, a peak sensitivity at high resolution 
of 7 mJy is more than 100 times the rms level. 
Figure 1 presents the low and high angular 
resolution Stokes $I$, $Q$, and $U$ images at the central ${\approx}1.''0$ 
region. 

The derived polarization intensity ($PI$) image is given by the quadrature
sum of the $Q$ and $U$ images, $PI=\sqrt{Q^2+U^2}$. 
The polarization angle in degrees is given by 
$E_{\rm{PA}}=(1/2)\times \arctan (U/Q)$.  
$PI$ and $E_{\rm PA}$ were calculated using a CASA task ``immath''. 
Debiasing level of 5$\sigma$ in Stokes $Q$ and $U$ was used in order to derive $PI$. 
The estimated error of the polarized intensity is,
$\Delta PI \approx \sqrt{\Delta Q^2+\Delta U^2 + (0.002*I)^2}$
which is the quadrature sum of the $Q$ and $U$ errors plus the nominal
polarization calibration error.  The estimated error of the
polarization angle in degrees is $\Delta E_{PA} \approx (1/2)\Delta
PI/PI$.  There is a lower limit of $\Delta E_{PA}$ of 
$\approx 1^\circ$ related
to the uncertainty of the position angle determination of the
polarization calibrator and antenna feed orientation. 
The degree of polarization is $D_{\rm{frac}}=PI/I$.  
Because of non-linearities in 
the parameters, only those pixels that contain both more than 5$\sigma$ signal level 
in the $PI$ maps and 3$\sigma$ signal level 
in the Stokes $I$ maps, were used for calculating $D_{\rm frac}$. 
In this paper, we use the term ``$E$-vector'' to refer to 
the observed polarization vector, and ``$B$-vector'' to refer 
to the observed polarization vector after rotating 90$^{\circ}$.

\subsection{The Molecular Line Observations}

In addition to the 1.1 mm polarization observations, 
we also performed observations in the CO (2--1) and SiO (4--3) line 
emission from the same science project in order 
to trace the molecular outflow associated with MMS 6. 
These observations were done separately, 
but the data were obtained in a similar period 
as the lower resolution polarization observations. 
Two SPWs were allocated to measure CO (2--1) and SiO (4--3) 
using the frequency division mode resulting 
in a the velocity resolution of $\approx$0.06 km s$^{-1}$. 
The standard data reduction script provided from 
the observatory was used to calibrate the data sets using CASA. 
The Briggs weighting with robust parameter of 0.5 was used 
for CLEAN binning with the velocity width to 3 km s$^{-1}$. 
The resulting synthesized beam sizes of the CO (2--1) 
and SiO (4--3) images are $0''.18{\times}0''.15$ (P.A.=$-16.5^{\circ}$) 
and $0''.22{\times}0''.17$ (P.A.=$-27.7^{\circ}$), respectively. 
The achieved rms noise levels of the CO (2--1) and SiO (4--3) images 
are 3.6 mJy beam$^{-1}$ km s$^{-1}$ and 
2.7 mJy beam$^{-1}$ km s$^{-1}$, respectively. 
The mean velocity maps (Figures 3 and 7) 
are produced from the data cube using 
a CASA task ``immoment'' with the clip levels of 3${\sigma}$ for CO (2--1) 
and 10${\sigma}$ for SiO (5--4). 
Details for the line data sets including both the analysis and interpretation 
will be presented in a forthcoming paper (Takahashi et al. 2019 in preparation).

\section{Results}
\subsection{Morphology and Polarization Vectors at Several 1000 AU Scale}

In Figure 2, we present the spatial distribution 
of the linearly polarized emission, Stokes $I$ emission, and $E$-vectors 
obtained from the 1.1 mm low angular resolution image. 
The image shows that the $PI$ peak is closely associated 
with the Stokes $I$ peak, while showing clumpy substructures. 
These substructures (central 200AU) will be described in $\S$ 3.2. 
In addition to the centrally concentrated substructures, 
we also detect extended emission both in Stokes $I$ and polarized emission. 
The emission is extended on 3000 AU scale and shows substructures. 
The most prominent feature is 
an ``arm-like structure'', detected in the northern part as 
denoted by the orange line in Figure 2. 
The arm-like structure is particularly clear in the $PI$ map, 
and also detected in the Stokes $I$ at a 5--15$\sigma$ level. 
The $E$-vectors are aligned with the minor axis 
of the ``arm-like structure'', and their orientations change smoothly along them. 
To the north of the arm-like structure, we find 
another faint component that is elongated north-south. 
This component was detected at a 5$\sigma$ level in the polarized emission, 
and a part of the structure was 
marginally detected in Stokes $I$ at a 3$\sigma$ level. 

In addition, the Stokes $I$ shows an extended faint 
(3--6$\sigma$) emission in the south-east part of the core. 
Unlike the northern arm-like structure, this component is 
not bright in the $PI$. 
Overall, $E$-vector orientations from the extended emission 
are aligned in the NW to SE direction, which is consistent 
with the large-scale orientation presented by \cite{2014ApJS..213...13H}. 

Figure 3 shows the Stokes $I$ image from a large squared region in Figure 2, 
and it is compared with the mean velocity map from the CO molecular outflow
\footnote{The CO outflow results will be 
presented in a separate paper 
(Takahashi et al. 2019 in prep.).}. 
The Stokes $I$ emission associated with the 
central compact component shows an elongated 
structure in the north-south direction 
with an extension of 3$''$ (1200 AU). 
The spatial distribution of 
the extended continuum emission shows enhancements 
on both edges of the CO molecular outflow. 
A depression in the Stokes $I$ emission 
is seen along the outflow, both in the north and the south. 
This implies that the interaction between the outflow 
and the surrounding material likely created the outflow cavity wall. 
The Stokes $I$ emission is not only detected from the area perpendicular 
to the molecular outflow where we normally expect infalling material to  
a pseudo disk, but is also detected in the direction of the outflow 
where material may be swept up by the outflow.

The $E$-vectors associated with this outflow interacting area, especially 
in the southern component (highlighted in yellow lines in Figure 2 
particularly at the south), show a different P.A. 
as compared to the global $E$-vector orientations, 
and seems to follow the interacting surface of the outflow.

\subsection{Morphology and polarization vectors in the central 200 AU}     

Figures 4(a) and 4(b) show the images of the central 200 AU region 
obtained in the low and high angular resolutions, respectively. 
The angular resolution of the image in Figure 4(b) of 
$0''.02{\times}0''.03$ is close to the highest angular resolution 
currently available with ALMA at 1.1 mm, and corresponds to a linear size scale of 
${\approx}8{\times}12$ AU at the distance of MMS 6. 
Our high angular resolution images show that the MMS 6 
continuum emission is very well resolved both in Stokes $I$ and in the polarized flux 
displaying a series of intricate structures.
The source size (FWHM) measured in Stokes $I$, 
is ${\approx}$165 synthesized beams. 
Furthermore, comparison of the high and low angular 
resolution Stokes $I$ images indicates that we are recovering 96\% of 
Stokes $I$ flux in the central 200 AU (i.e., central $0''.5$, 
the region denoted in the dashed circle in Figure 1a). 
This shows that nearly little or none of the emission 
in the central region is contained in a large, resolved-out, component.

Both the Stokes $I$ and $PI$ emissions 
show a peak around the center. However, the $PI$ peak is offset 
to the south-west (SW) by $0''.05$ ($\approx$20 AU) with 
respect to the Stokes $I$ peak. Hereafter we will refer to 
this concentrated $PI$ structure 
as the ``centrally concentrated component'' (Figure 4c).  
The emissions of both Stokes $I$ and $PI$ are 
elongated along the NE to SW direction with P.A.${\approx}45^{\circ}$. 
The associated $E$-vectors are aligned to the minor axis of 
the elongated structure (P.A.$\approx$135$^{\circ}$). 
Around this central structure, there are three more components 
called ``NW'', ``SE'' and ``West'' components (identified in Figure 4c), 
each displaying substructures, that, together, 
define an approximate ring surrounding the central component. 
Hereafter, we call this the ``ring-like structure'' (Figure 4c). 
The substructures seen within the ``ring-like structure'', 
are particularly clear within the SE component, with  
a size scale as small as the synthesized beam size ($\approx$10 AU). 

The $E$-vectors associated with the NW and SE components 
are more or less azimuthally aligned (Figure 4b). 
The position angle of the $E$-vectors changes in the ``ring-like structure''; 
i.e., the peaks of the P.A. distribution are 25$^{\circ}$ and 40$^{\circ}$ 
for the SE and NW components, respectively, while the peak of the P.A. 
distribution is 135$^{\circ}$ at the ``centrally concentrated component''. 
Significant changes in the P.A. 
of 90$^{\circ}$ -- 105$^{\circ}$, occur in the central $0''.5$ region. 
The P.A. within the NW and SE components 
varies slightly with respect to the overall $E$-vector orientation. 
This implies that our data not only reveal the bulk structure, 
but also the small scale $E$-vector changes, 
which may be related to small scale variations in the magnetic field, 
due to local structural changes, dust property changes, or dynamics. 

Finally, there is a clear polarization gap between 
the ``centrally concentrated component'' and the ``ring-like structure''. 
The width of the gap is significantly larger than 
the beam size, and is therefore not caused by the beam dilution. 
This gap implies that either the $PI$ is intrinsically low 
or that the $E$-vector orientations originating 
from the centrally concentrated component and the ring-like structure 
are approximately orthogonal along the line of sight.

\subsection{Polarization Degree}

In Figures 5(a) and (b), we present the degree of the polarization, $D_{\rm frac}$, 
as derived from the low angular resolution image. 
We find that $D_{\rm frac}$ in the arm-like structure 
shown in Figure 5(a), has a relatively high value of 15-20\%. 
For the central region in Figure 5(b), $D_{\rm frac}$ 
is less than 3\% for most of the area, 
but there are local peaks with $D_{\rm frac}$ up to ${\approx}$8\%. 
The locations of the peaks in $D_{\rm{frac}}$ coincide with the local maxima of $PI$. 

The distribution of $D_{\rm frac}$ obtained with the 
high angular resolution image is presented in Figure 5(c) and
the comparison of this central region between the low and high 
resolution images in $I$, $Q$ and $U$ is shown in Figure 1. 
The internal structures within the central 200 AU are 
spatially resolved and show the following components: 
(i) the ``centrally concentrated component'', 
showing the peak in $PI$ with $D_{\rm frac}$ of 11\%, and 
(ii) the partial ``ring-like structure'' with $D_{\rm frac}$ up to 19\%. 
$D_{\rm frac}$ is particularly high within the SE component. 
The locations of all the local maxima of $D_{\rm frac}$,  
coincide well with the local peaks in $PI$. 

Considering the central $0''.5$ region 
(denoted by the dashed circle in Figure 1a), 
there is no significant missing Stokes $I$ flux. 
Therefore the derived $D_{\rm frac}$ 
is not overestimated and shows 
most likely the intrinsic value. In contrast, for 
the central $2''.0$, comparison with low resolution maps suggests 
that about 25\% is missing in the Stokes $I$ emission. 
The relatively high $D_{\rm frac}$ (7--25\%) 
derived for the extended structure, in the NE side 
with respect to the center in Figure 5(b), may be somewhat overestimated. 

Finally, note that a comparison between the low 
and high resolution images presented in Figure 5(b) and 5(c) 
clearly shows that $D_{\rm frac}$ becomes higher 
for the high angular resolution image toward local substructures.  
Those substructures contain highly organized $E$-vectors 
as small as a few $\times$10 AU scale, 
while they are diluted by the beam and show considerably 
less $D_{\rm frac}$ in the low resolution map.

\subsection{Origin and Physical Properties of the 1.1mm Continuum Emission}

The total flux of the 1.1 mm continuum emission is measured to be 0.9 Jy 
using the area where the SNR is greater than 5 from 
the high angular resolution image. 
This flux includes the possible contribution 
of the free-free emission from the ionized 
jet from the protostar. 
The free-free emission can be extrapolated from the centimeter observations, 
with an assumption of the frequency dependence of $F({\nu}){\propto}{\nu}^{0.6}$ 
\citep{1998AJ....116.2953A,1986ApJ...304..713R}. 
\citet{2009ApJ...704.1459T} measured a 3$\sigma$ upper limit of 0.09 mJy 
in the 3.6 cm continuum band. Adopting this number as the upper flux limit 
attributed to the free-free jet, the free-free contribution at the 1.1 mm band 
is at most 0.7 mJy, which is $\lessapprox$0.8\% 
of the total Stoke $I$ flux
\footnote{No significant time variation 
in the free-free emission is assumed.}. 
This indicates that the thermal dust emission is dominating 
in the 1.1 mm continuum band, and that the contribution from 
the free-free emission of the ionized jet 
is within the measurement errors. 

In order to characterize the internal structures and their physical 
parameters from 
the Stokes $I$ high angular resolution image, 
we used a two-dimensional Gaussian fit 
with multiple components. 
The fitting results are summarized in Table 2 
and show that with three Gaussian components; 
(i) an extended component, 
(ii) an elongated component, 
and (iii) a compact component, 
The observed structures are well reproduced. 
The residual level of the fitting result is less than 1.5\% 
with respect to the observed peak flux 
(i.e., the residual level is less than  SNR${\leq}1.7$). 
Note that the fitting region 
includes the NW, SE, West components. 
However, those components are only bright in the $PI$ image. 
Hence Stokes $I$ fitting results are not affected by those 
substructures, but rather based on the total material distribution.

The extended component shows structure sizes of 
${\approx}0''.3$ ($\approx$120 AU) in FWHM. 
This size is comparable to the maximum recoverable size of the experiments. 
For a component that is more extended, flux will be missed 
by the lack of short antenna spacings. 
The elongated component has a fitted size 
of $0''.3{\times}0''.05$ (120 AU $\times$ 20 AU), 
with an aspect ratio of 5.8 with P.A.=39$^{\circ}$. 
Finally, a compact component of $0''.05{\times}0''.01$ (20 AU $\times$ 4 AU) 
is necessary in order to explain additional flux seen at the center. 
This compact component is barely resolved with respect to the synthesized beam. 

The measured 1.1 mm peak intensity of 7 mJy beam$^{-1}$ 
corresponds to a brightness temperature of 192 K. 
Gas temperatures of $\gtrsim$100 K are expected at 
the radius of $\approx$10 AU from the protostellar source 
(e.g., \citealt{2003ApJ...583..322N}). 
Assuming that the measured peak flux mainly comes 
from the circumstellar materials 
associated with the central protostar, 
the estimated high brightness temperature 
of the dust can be compared to the expected 
gas temperature at the radius of 10 AU. 
We conclude that the observed dust emission 
may become optically thick at radii $\lesssim$10 AU. 

On the other hand, the measured peak flux may be attributed to 
the multiple components integrated along the line of sight.
The mean brightness temperatures estimated from the fitted total fluxes  
listed in Table 2 are 25 K (extended component), 
27 K (elongated component), and 36 K 
(compact component). 
These are considered as the lower limits 
of the dust temperatures. 
From a theoretical approach, \citet{2017ApJ...835L..11T} 
presented the temperature distribution 
around the center region of the protostellar core 
using radiative transfer calculations. 
They found that the observed temperature 
integrated along the line of sight 
is $\approx$20 K in the mid-plane of the pseudo 
disk at $r{\lesssim}200$ AU due to the optical depth effect. 
The fitted three components also have similar size structures 
(${\approx}0''.3$ or $\approx$120 AU), 
thus likely trace emission from the pseudo disk. 
The observed brightness temperatures are lower 
than predicted by \citet{2017ApJ...835L..11T}. 
This implies that MMS 6 also has an optically-thick 
pseudo disk with a relatively low temperature.

Assuming the respective mean dust temperatures of 25, 27, and 36 K for the extended, 
elongated, and compact components, an optical depth of ${\tau}{\approx}1$, 
a gas-to-dust ratio of 100, 
we have estimated the lower limits of the column density ($N_{\rm{H_2}}$), 
the gas mass ($M_{\rm{H_2}}$), 
the number density ($n_{\rm{H_2}}$) as listed in Table 3. 
Here, the brightness temperatures estimated 
for the elongated and compact components  
are adopted as the dust temperature. 
A spherical geometry 
with the geometric mean of the major and minor axes 
of the source size is assumed in order to estimate the number densities. 

The dust temperature for the extended component 
is assumed from previous multi-wavelength observations of 
the large scale structure \citep{2016A&A...588A..30S}. 
For those estimates, we adopted a dust emissivity index 
of $\beta$=0.93 from \citet{2009ApJ...704.1459T}, 
and a dust absorption coefficient of 
${\kappa_{\lambda_{\rm{(obs)}}}}$=0.037 cm$^2$ g$^{-1}$ 
(${\lambda_{\rm{obs}}}$/400$\mu$m)$^{-{\beta}}$ by \citet{1982ApJ...252L..11K} 
where $\lambda_{\rm{obs}}$ the observed wavelength of 1.1 mm 
with ${\kappa_{\lambda_{\rm{(1.1mm)}}}}$=0.014 cm$^2$ g$^{-1}$.
Note that, for the large-scale emission ($\gtrsim$0.05 pc), 
$T_{\rm{dust}}$=25 K and $\beta$=1.4 were estimated toward MMS 6 
from $Herschel$ and GISMO/IRAM 30m data sets \citep{2016A&A...588A..30S}. 
Adopting their value of $\beta$, the estimated dust 
masses increase by a factor of 1.7. 
Another factor that affects the estimates 
is the mass absorption coefficient. 
\cite{2016A&A...588A..30S} adopted $\kappa_{\nu_{\rm{(obs)}}}$=0.1 cm$^2$ g$^{-1}$ 
($\nu_{\rm{obs}}$/1.0 THz)$^{\beta}$ based on $Herscel$ results by \citet{2010A&A...518L.102A}. 
Adopting this number, $\kappa_{\rm {1.1mm}}$ increases by a factor of 2.3, 
resulting in a corresponding decrease in the estimated dust masses. 

The fitted source size of the extended component of ${\approx}0''.3$ is consistent with 
the value reported from previous lower resolution 
SMA observations of the dust continuum emission 
at 850$\mu$m \citep{2012ApJ...752...10T}.   
The estimated total mass in this paper, $M_{\rm{H_2}}{\approx}1$ $M_{\odot}$, 
is about three times larger than those in \citet{2012ApJ...752...10T} 
because we have used in this paper a dust temperature that is 
lower by a factor of 2. 
Assuming the same dust temperature as used in \citet{2012ApJ...752...10T},
the derived physical parameters ($M_{\rm{H_2}}$, $N_{\rm{H_2}}$, and $n_{\rm{H_2}}$) 
in this paper would be more consistent with previous results 
(e.g., the estimated mass would be agreement within a factor of 1.3).

\section{Discussion}

As described in $\S$3, the $E$-vectors 
derived from our ALMA 1.1 mm dust continuum emission data 
show totally different field patterns as compared with 
previous interferometric studies 
such as obtained with BIMA by \citet{2005ApJ...626..959M} 
or with CARMA by \citet{2014ApJS..213...13H}. 
One obvious factor is a significant difference in beam sizes, 
which is a factor of $\approx$10000 
in terms of the beam surface area, between the previous 
experiments and our ALMA high angular resolution data. 
It is natural to consider that the polarized 
emissions from the two different spatial scales trace totally 
different physical structures and phenomena inherent to each spatial scale. 
Comparison of these different angular-size structures is important 
to understand how the magnetic field morphology is connected 
from a pseudo disk (${\approx}10^3$--$10^4$ AU) to 
a circumstellar disk ($\lesssim$ a few $\times$100 AU). 

In addition, the origin of the linearly polarized dust 
continuum emission in the vicinity of protostars 
is not completely understood yet. It has been 
suggested recently that the linearly polarized 
(sub)millimeter dust emission may not be always 
associated with the magnetic field 
(e.g., \citealt{2018ApJ...855...92C,2018ApJ...861...91H,2018ApJ...856L..27G,2018ApJ...859..165S}). 
In \S4.1, we summarize other possible mechanisms to 
align dust grains and/or to produce polarized emission, and, in \S4.2, 
we compare the observed ALMA data with the various models. 
Finally, in \S4.3, we further discuss the magnetic field morphology 
as inferred from the polarized emission 
and compare it with MHD simulations.

\subsection{Potential Origins of the Polarized Dust Emission}

Figure 6 shows three possible mechanisms 
to produce polarized emission in the millimeter 
and submillimeter wavelengths.
The first mechanism to produce the polarized dust 
emission is from magnetically aligned dust grains as 
proposed originally by \citet{1949Sci...109..461S} -- see also 
\citet{1951ApJ...114..206D}, \citet{1988QJRAS..29..327H}, 
and \citet{2007JQSRT.106..225L}. 
Spinning dust grains interact with the magnetic fields 
and align their major axes perpendicular to the magnetic field. 
Therefore, thermal radiation from these grains produce polarized 
emission with the $E$-vector aligned perpendicular to the magnetic field 
as presented in Figure 6(A) (e.g., \citealt{2007JQSRT.106..225L}). 
This is what has been mainly detected 
on the size scales of clouds, cores, and envelopes 
(e.g., \citealt{1998ApJ...502L..75R,2009ApJ...707..921R,2003ApJ...598..392L,
2014ApJS..213...13H,2017ApJ...847...92H,2006Sci...313..812G,2009Sci...324.1408G,
2013ApJ...772...69G,2014ApJ...792..116Z,2013ApJ...763..135T,
2018ApJ...855...39K,
2018arXiv180507348K,2018ApJ...854...56L}). 

The second mechanism to produce millimeter wavelength polarized dust 
emission is self-scattering of the dust grains as 
proposed by \citet{2015ApJ...809...78K} and \citet{2016MNRAS.460.4109Y}. 
Even without any alignment of the grains in the disk, polarized emission 
is expected when the dust grains scatter an anisotropic
radiation field. If the radiation energy flux is dominated in the azimuthal
direction rather than in the radial direction, the self-scattering 
is expected to produce net polarization in the radial direction 
as shown in Figure 6(B) by the polarization from a dust ring. 
\citet{2015ApJ...809...78K} predicted that the scattering 
at millimeter wavelengths is 
efficient if the dust grains are as large as 100 $\mu$m. 
In this model, a polarization degree of a few percent is expected for 
assuming spherical dust grains. 
This mechanism explains well some of the recent ALMA polarization results 
obtained toward {protoplanetary disks or circumstellar disks including: 
HL Tau \citep{2017ApJ...844L...5K,2017ApJ...851...55S}; 
HD 14527 \citep{2016ApJ...831L..12K}; 
IM Lup \citep{2018ApJ...860...82H}; 
DG Tau \citep{2018ApJ...865L..12B}; 
HD 163296 \citep{2019MNRAS.482L..29D}; 
VLA 1623 \citep{2018ApJ...859..165S,2018ApJ...861...91H}; and 
GGD 27 MMS 1 \citep{2018ApJ...856L..27G}, as well as 
Class 0/I sources in the Perseus 
Molecular Cloud \citep{2018ApJ...855...92C}.}

The third mechanism comes from dust grain alignment due to the anisotropic radiation 
as proposed by \citet{2017ApJ...839...56T}. 
Near the star, the dust alignment solely due to the anisotropic radiation 
is expected in the direction determined by the radiation flux. 
The dust grains are expected to be aligned with their minor axis 
parallel to the radiation direction as illustrated in Figure 6(C). 
The polarization degree depends mainly on the maximum grain size, 
the shape of the grains, and the intrinsic alignment efficiency of the grains.

\subsection{Comparison of Each Model}

The ALMA polarization images presented in Figures 2 and 4 highlights 
the complex substructures within 
the protostellar core, MMS 6. Therefore, it is possible that 
the polarized emission may be different by regions. 
In this section, we discuss possible origins of the dust polarization 
by regions using the proposed three theoretical models. 

\subsubsection{Origin of the Polarized Dust Emission Scales of 
a few $\times$1000 AU}

As described in \S 3.1, the organized $E$-vectors 
with P.A.${\approx}135^{\circ}$ is detected at low angular 
resolution image presented in Figure 2. 
The $E$-vector orientations are consistent with those previously 
measured in the filament, the core (a few $\times$0.1 pc),  
and the envelope size scales (a few $\times$1000 AU) as reported by \cite{2000ApJ...531..868M,2001ApJ...562..400M,2005ApJ...626..959M,2010ApJ...716..893P,2014ApJS..213...13H}. 
The mechanism to align the dust grains in those extended 
structures has been widely accepted to be the magnetic field. 
Hence, the origin of the extended polarized emission 
measured by the ALMA low angular resolution image appears 
consistent with a magnetic alignment of dust grains.

The $E$-vectors detected in the ``arm-like structure'' (Figure 2), 
are aligned perpendicular to the elongated structure. 
Assuming that the mechanism producing the polarized emission is 
the magnetic alignment of dust grains (Figure 6A), the magnetic fields 
then run along the ``arm-like structure'' as denoted in Figure 5(a). 
However, without investigating the gas dynamics, it is difficult to conclude 
what the origin of the ``arm-like structure'' is. 
This structure could be explained by a physical 
arm or just asymmetrical substructures within the pseudo disk. 
Similarly, extended substructures likely associated with 
the envelope are also detected toward Class 0/I sources, for example  
Per-emb-11, Per-emb-29, Per-emb-2, and Per-emb-5, 
in the Perseus Molecular Cloud by \cite{2018ApJ...855...92C}. 
They are also bright in the polarized emission and 
$E$-vectors are aligned to the minor axis of those extended substructures. 
\cite{2018ApJ...855...92C} interpreted this 
as a part of the hourglass morphology expected 
if the polarization traces magnetic 
field dragged by the accreting material.  
Their morphology, size scale, and $E$-vector orientations 
are similar to those observed in the ``arm-like structure'' in MMS 6.
In any case, the magnetic field does not 
deter gas infalling motions toward the central protostar 
through this structure. 
In \S 4.3, we will make further comparisons of 
the magnetic field models using a MHD simulation result.

\subsubsection{Origin of the Polarized Dust Emission within the Central 200 AU}

Our ALMA high angular resolution polarization image 
and the polarization $E$-vectors within the central 
200 AU show the two major components of 
``centrally concentrated component'' 
and ``ring-like structure'' (Figure 4). 
We also identify the following characteristics: 
(i) there are very different spatial distributions between the Stokes $I$ and the $PI$, 
(ii) there is a clear positional offset between the peak positions of the Stokes $I$ and the $PI$, 
(iii) a clear ring-like gap that shows no detection of the polarized emission, 
(iv) and a significant change in the P.A. of the polarization 
$E$-vector occurs across the gap, 
and finally (v) we measure a high polarization degree 
$\gtrsim$10\%) within the central 200 AU. 
Each of these characteristics may be crucial 
in identifying the polarization mechanisms at play, and, hereafter 
we examine which of the five characteristics can be explained 
by the proposed polarization models. 
Table 4 summarizes how each of the characteristics 
could fit in each proposed polarization mechanism. 

{\bf First, we consider the dust grain alignment 
due to the anisotropic radiation (Figure 6C).}  
The azimuthally aligned polarized emission is expected for this case. 
The ``ring-like structure'', shown in MMS 6, shows 
a similar azimuthal pattern but it does not trace 
back precisely to the center. 
In particular, the NW component is pointed to the NE 
of the Stokes $I$ peak position.  
In addition, the $E$-vectors associated with 
the ``centrally concentrated component'' 
are aligned with the minor axis 
of the disk-like structure, but not pointing 
toward the protostar, which is different from 
the predictions of the anisotropic radiation model. 
Our observational results, therefore, 
cannot unambiguously support this scenario.

{\bf Second, we consider the polarization $E$-vectors originated from  
the self-scattering of the dust (Figure 6B).} 
The observed polarization $E$-vector is aligned with 
the minor axis of the ``centrally concentrated component''. 
This is not inconsistent with the prediction of the $E$-vector 
orientation for a tilted disk 
\citep{2015ApJ...809...78K,2016MNRAS.460.4109Y,2017MNRAS.472..373Y}. 
The observed positional offset between the Stokes $I$ emission 
and the polarized emission can be explained by the optical depth effect, 
which is presented by \cite{2017MNRAS.472..373Y}.  
Their model considers a finite optical depth and a finite thickness of the emitting 
and scattering dust grains within the disk, and then calculate the optical 
depth effect analytically.
Their calculation showed that the absorption optical depth remains well below
unity. However, the scattering optical depth becomes larger than unity. 
This produces an asymmetry of the polarized emission. 
Assuming that the disk-like structure is perpendicular to 
the blue and red shifted outflow gas presented in Figure 3, 
the near-side and far-side of the disk are 
then on the SE and NW sides, respectively. 
Our results show that the peak position of the polarized emission 
is shifted to the near-side of the disk (i.e., south-east) with respect to 
the peak position of the Stokes $I$ emission. 
This is consistent with what \citet{2017MNRAS.472..373Y} predicted. 

Recent ALMA results of two embedded protostellar sources, 
GGD27 MMS 1 (an early B-type protostar; \citealt{2018ApJ...856L..27G}) 
and VLA 1623 in $\rho$ Ophiuchus (a proto-binary Class 0 source; 
\citealt{2018ApJ...859..165S,2018ApJ...861...91H}), show similar 
features that the one we observe in MMS 6. 
Both GGD27 MMS 1 and VLA 1623 show 
$\approx$200 AU scale circumstellar disks, 
both of which are likely optically thick. 
Their $E$-vector orientations show 
the azimuthal ring-like pattern 
at the outer part of the disk, 
while the $E$-vectors at the inner disk 
are aligned to the minor axis of the disk 
(i.e., aligned to the minor axis of the elongated 
continuum structure). A positional offset between 
the Stokes $I$ and $PI$ was reported for these 
two sources, and this can also be explained 
by the optical depth effect \citep{2017MNRAS.472..373Y}. 
These features are very similar to what we see in MMS 6. 

On the other hand, there are some significant differences 
between MMS 6 and other objects. 
First, the ``centrally concentrated structure'' 
(i.e., elongated continuum structure at the center) 
detected in MMS 6 is not extending 
in the perpendicular direction to the axis of the outflow, 
hence it is not clear whether the structure is actually a disk. 
This is different from structures seen in GGD27 MMS 1 and VLA 1623, 
both of which show that the disks are clearly extending in the 
 perpendicular direction to the axis of the outflow. 
Second, the polarized emission should be 
correlated more or less with the Stokes $I$ emission 
if the origin of the polarized emission is dominated 
by the self-scattering because the amount of polarized emission  
correlates with the total flux. 
This is not the case for MMS 6. 
Third, and most importantly, the measured polarization degree in MMS 6 
is very high compared to any other sources within the central 200 AU scale. 
For example, in both GGD27 MMS 1 and VLA 1623, 
the polarization degree is measured to be less than 3\% at the inner disk, 
while in the case of MMS 6 shows the polarization degree as high as 11\% 
at the central peak and over 20\% at the NE part of 
the ``ring-like structure''. 

Furthermore, another recent ALMA study by \cite{2018ApJ...855...92C} 
of 10 young Class 0/I protostars shows that the polarization degree 
is relatively high (${\gtrsim}5\%$)  
in the envelope, while low in the inner disk (${\lesssim}1\%$). 
They suggested that self-scattering is dominant in the inner disk, 
while the magnetically aligned dust is likely dominant in the envelopes. 
In contrast, MMS 6 shows consistently high polarization degrees 
in the central 200 AU, both for the 
``centrally concentrated structure'' and the ``ring-like structure''. 
Note that our MMS 6 image has a 10 AU resolution, while the study by \cite{2018ApJ...855...92C} has a resolution of only $\approx$80 AU.  
Alternative possible explanations of the very different polarization 
degree observed within the central 200 AU 
may be caused by the beam dilution effect as pointed out in \S3.3. 

In summary, the self-scattering model can explain the polarization 
pattern and positional offset between the Stokes $I$ and the $PI$, 
detected at the ``centrally concentrated component''. 
This is consistent with other protostellar sources such as; GGD27 MMS 1 and VLA 1623. 
However, the self-scattering model cannot reproduce 
the high polarization degree observed in MMS 6.  
In order to further explore the possibility of 
the self-scattering effect for MMS 6,  
multi-wavelength dust polarization experiments are crucial. 
If self-scattering is a dominant mechanism, we expect changes 
in the polarization degree as a function 
of the wavelength \citep{2015ApJ...809...78K}. 

{\bf Finally, as a third model, we consider the magnetically aligned 
dust grain model (Figure 6A).} 
In this model, the direction of the (uniform) interstellar magnetic field is 
given by the direction of the $B$-vector. 
Thus, the magnetic field projected onto the celestial plane is obtained as 
the perpendicular direction to the observed $E$-vectors as presented in Figure 5(c). 
In the ``centrally concentrated component'', the interstellar magnetic 
field seems to run parallel to the major axis. 
The P.A. is measured to be 45$^{\circ}$, 
which is consistent with the orientation of the large-scale magnetic field 
\citep{2000ApJ...531..868M,2001ApJ...562..400M,2005ApJ...626..959M,
2010ApJ...716..893P,2014ApJS..213...13H}. 

On the other hand, the magnetic field vectors 
located on the ``ring-like structure'', especially at the SE and NW components, 
show the $E$-vectors (P.A.=115--130$^{\circ}$) 
almost perpendicular to those associated 
with the ``centrally concentrated components'' .
One possible way to explain the observed field configuration 
is a toroidal wrapping of the magnetic field lines. 
\cite{2011PASJ...63..147T} modeled magnetically driven molecular 
outflows through MHD and radiative transfer simulations. 
The results suggest that the toroidal magnetic field is dominant 
on $\lesssim$400 AU (i.e., size scale of pseudo disks) and 
the projected magnetic field vectors change significantly 
depending on their viewing angle. 
According to Figure 5 of \cite{2011PASJ...63..147T}, 
ALMA results share some characteristic structures 
expected from the simulation results when choosing 
the inclined models (i.e., the viewing angle of 
45$^{\circ}$-60$^{\circ}$ in this model). 
First, the magnetic field structures at the 200 AU scale 
do not seem to follow the orientation of the large scale 
outflow, but rather are associated with the toroidal 
component of the pseudo disk. 
Second, the brightness of the polarized emission within the 
toroidal component is not uniform depending on the azimuthal angle. 
The ``ring-like structure'' 
observed in MMS 6 somewhat looks similar to 
those presented in \cite{2011PASJ...63..147T}. 
Moreover, the relation between the magnetic 
field orientation and the core rotational axis (outflow axis) 
also adds another important factor in determining 
the magnetic field orientations \citep{2012ApJ...761...40K}. 
This will be discussed further in \S4.3.

Among the three proposed scenarios, 
the magnetic field model seems to explain 
the overall distribution of both 
the Stokes $I$ and polarized emission, 
not only the displacement of the peaks of Stokes $I$ 
and polarized emission, but also the ``ring-like structures'' 
and depolarization gaps. 
The Stokes $I$ emissions correspond to the spatial 
distribution of the column density combined 
with the dust temperature, while the polarized 
emissions correspond to the locations where 
the $E$-vectors, and therefore the magnetic fields, 
are most organized, and not necessary to correlate 
with the Stokes $I$ emission. 
The observed high polarization degree, 
that cannot be reproduced by the self-scattering 
mechanism, can also be explained by 
the magnetically aligned dust model. 

In addition to the listed individual mechanisms, 
the importance of a hybrid model combining them, 
and the change of the dust alignment mechanisms across the wavelengths, 
has been pointed out from the ALMA observations of HL Tau \citep{2017ApJ...844L...5K,2017ApJ...851...55S}. 
Their results suggest that 
the self-scattering mechanism dominates at 870 $\mu$m, 
a hybrid model of the self-scattering and 
the dust alignment due to the anisotropic radiation 
explains the results obtained at 1.3 mm, 
and the anisotropic radiation mechanism dominates at 3.1 mm. 
Multi-wavelength polarization experiments will therefore 
be essential in order to distinguish the origins of the polarized emission 
and to disentangle the multiple mechanisms at play.

Moreover, we would like to note that a strong wind or jet 
might mechanically align the dust grains. 
\cite{1952MNRAS.112..215G} discussed dynamical interaction 
between gas and elongated dust particles. 
In this model, the major axis of dust grains prefers to 
align parallel to the supersonic gaseous flow, 
hence the $E$-vectors should be observed along the outflow axis. 
Figure 7 presents a zoomed image of the SiO collimated outflow
\footnote{SiO data will be presented in a separated paper 
(Takahashi et al. 2019, in prep.)} 
overlaid with the polarized emission. 
The observed $E$-vectors are aligned 
perpendicular to the cavity of the outflow, particularly 
in the southern lobe
\footnote{Majority of the $E$-vectors on the northern lobe seems to 
trace the large-scale magnetic fields (P.A.$\approx$45$^{\circ}$ in Figure 7b), 
which likely originated from the foreground fields associated with 
the cloud or envelope.}. 
The observational result does not support the scenario 
expected by \cite{1952MNRAS.112..215G}, while the result 
is more consistent with a scenario proposed by 
\cite{2007ApJ...669L..77L}, involving an alternative mechanical 
alignment of helical dust grains. 
In this scenario, the helical dust grains align 
with the minor axis being parallel to the gaseous flow 
(both supersonic and subsonic cases). 
This results in $E$-vector that are perpendicular to the gas outflow. 
However, the magnetic alignment also predicts $E$-vectors 
perpendicular to the gas outflow if the magnetic field is parallel to the gas velocity. 
Thus, the mechanical alignment proposed by \cite{2007ApJ...669L..77L} 
and the magnetic alignment predict the 
same polarization pattern for the outflow region. 
Recent ALMA dust polarization observations reported 
similar $E$-vector configurations, likely tracing the outflow 
cavity in two other Class 0 sources, namely: B335 \citep{2018MNRAS.tmp..552M} 
and L1157 \citep{2018arXiv180507348K}.  

Finally, it is interesting to note that the polarized emission 
peaks seem to avoid to overlap with the SiO collimated outflow. 
The outflow may have impacted the ``ring-like structure'', 
pushing aside the gas. Alternatively, the outflow and the ring-like 
structure are not in the same plane, but are projected along the line of sight.

\subsection{Comparisons with the MHD simulations}

High sensitivity and high angular resolution 
at millimeter and submillimeter wavelengths of the polarized dust emission
became available with ALMA, enabling us to make direct comparisons 
between high angular resolution images and detailed theoretical models.  
In this section, we present comparison using our ALMA data 
MMS 6 with MHD simulations, and we discuss how 
the observational results could be further interpreted 
with a magnetic field model (i.e., the third model discussed in \S4.2.2). 

In order to take account of the viewing angle of the system, 
the relation between the magnetic field orientation 
and the core rotational axis (outflow axis), 
and finally compare the model results with our ALMA data, 
we executed a three dimensional non-ideal MHD simulation 
according to the following steps. 
As the initial state of the prestellar core, we assumed 
a Bonnor-Ebert sphere with a central density of 
2 $\times$ 10$^5$ cm$^{-3}$ and an isothermal temperature of 10 K. 
A uniform magnetic field ($B_0$=26 $\mu$G) and rigid rotation 
($\Omega_0$ = 1.0 $\times$ 10$^{-13}$ s$^{-1}$) are imposed, 
in which the magnetic field direction is inclined from 
the rotation axis by 45$^{\circ}$. 
We use the Cartesian coordinate, in which we choose the direction 
of the $z$-axis to coincide with the direction of 
the initial magnetic field, 
i.e., $\vec{B}_0=(0,0,B_0)$. 
The rotation axis is in the $x-z$ plane as 
$\vec{\Omega_0}=\Omega_0(1/\sqrt{2},0,1/\sqrt{2})$.
Here, the offset between the magnetic field axis and rotation axis is 
determined from the measured magnetic field direction 
\citep{2014ApJS..213...13H} and the outflow axis \citep{2012ApJ...745L..10T}. 
The mass and radius of the prestellar core are $M_{\rm{core}}$=8.7 $M_{\odot}$, 
and $R_{\rm{core}}$ = 2.5 $\times$ 10$^4$ AU, respectively. 
Using the nested grid method (for details, see \citealt{2005MNRAS.362..369M}), 
we calculated the core evolution from the protostellar run-away collapse phase 
to the protostellar phase, during which the outflow is launched, 
and the outflow reaches $\sim$ 5000 AU. 
The cell size of the finest grid is 0.2 AU. 
The details of the MHD simulation are described in \cite{2013MNRAS.431.1719M}. 
Note that, in this simulation, we used a sink cell technique, 
in which the central region was masked by the sink cell. 
Thus, we cannot resolve the central region or protostar itself.

The column density (reflects Stokes $I$), 
the polarized emission, and the magnetic field 
vectors are calculated, 
and integrated along the line of sight in order 
to make a projected polarization map \citep{2011PASJ...63..147T}. 
Optically thin condition was assumed when making this projection. 
A viewing angle of $i$=60$^{\circ}$ was used 
\footnote{The inclination angle $i$ is the angle between 
the direction of the initial magnetic field  
and the line of sight. $i$=0$^{\circ}$ is defined as the $z$-direction.} 
This viewing angle was chosen so that the blue and red shifted outflow lobes 
do not overlap in agreement with previous outflow observations 
\citep{2012ApJ...745L..10T}. 
Also the inclination angle of 60$^{\circ}$ appears to be a better fit 
to explain our observed results as presented in Figure 5(c). 

Figure 8 presents the MHD simulation result 
obtained for MMS 6 with $i=60^{\circ}$. 
Figure 9(a) and 9(b) show the expected polarization maps 
plotted for two different size scales. 
We found that our ALMA images presented in Figure 5(a) and 5(c) share 
some characteristic structures predicted from the MHD simulation. 
First, the simulation results show that arms 
on the size scale of $\gtrsim$500 AU are connected 
to the central component (Figure 9a). 
The magnetic field structures ($B$-vectors) within 
the arms are aligned along the major axis. 
The spiral arms are seen within the pseudo disk, 
which is produced owing to the magnetic field and 
the core rotation. 
The material in the arms accretes towards 
the central region through the spiral arms. 
Our ALMA data presented in Figure 5(a), 
show that the ``arm-like structure'' extend to the 2000 AU scale.  
The $B$-vector orientations are also aligned 
along the major axis of the structure, and 
the ``arm-like structure'' is connected to 
the central component from the western side. 

Second, within the central 100 AU, the MHD simulation result 
shows a ring-like structure that produces a relatively high 
$PI$(Figure 9b). We see local peaks within 
the ring-like structure due to the effect of integrating polarized 
emission along the line of sight. The magnetic field vectors in the ring-like structure 
are aligned more or less along the minor axis of the structure with some fluctuation. 
In this simulation, while the $PI$ 
shows a ring-like emission distribution, the column density 
(reflects Stokes $I$) increases toward the center.  
Moreover, within the ring-like structure, we can see $PI$ 
variation (local peaks). $B$-vector orientations in the ``ring-like structure'' 
are also aligned along the minor axis of the structure as seen in Figure 4(c).
In addition, the ALMA observations show a depolarization within this structure. 
These qualitative features seen in the MHD simulation 
are all similar to what we see in the observational results 
(Figures 4b and 5c). 
The difference is a bright compact component, 
which is only seen in the ALMA data  
and is not reproduced in the MHD simulation. 
Because we calculated 
the MHD simulation using a sink technique, 
the current simulations neither reproduce 
the protostar itself nor include the radiation 
field from the central protostar. 
This underestimate emission from 
the circumstellar disk at the most internal region. 
At this moment, it is not obvious how these 
effects are reflected in the polarized emission, 
magnetic field, and polarization 
degree for the innermost region, 
where we see the ``centrally concentrated 
component'' from the ALMA data. 

In summary, by adjusting the viewing angle 
around the outflow driving region, 
the present MHD simulation can qualitatively explain the characteristic features 
of the ALMA observations such as the $B$-vector orientations 
both in the large (1000 AU scale) 
and small (within a few 100 AU) scales. 
The magnetic field model also explains 
differences in the spatial distribution 
between the Stokes $I$ and the $PI$.

\section{Summary and Future Prospects}

Using ALMA, we performed full polarization high angular resolution observations of the dust continuum emission at 1.1 mm toward a very young intermediate mass protostellar source, MMS 6/ OMC-3. 
We have achieved the spatial angular resolution of $0''.03{\times}0''.02$ 
corresponding to a linear size scale of $\approx$10 AU at the distance to the source. 
The main findings of the study are as follows. 

\begin{enumerate}

\item	Our high angular resolution image 
shows the complexity of the total and $PI$ 
within 200 AU of the center. 
While Stokes $I$ emission shows a single peak toward the center, 
the dust polarization map shows two different components 
which are characterized as the ``centrally concentrated component'' and the ``ring-like structure'' with a clear depolarization gap between the two components.  
Stokes $I$ emission and $PI$ also show a clear positional offset. 

\item	The detected $E$-vectors are spatially resolved in the central 200 AU 
and clearly show organized structures. 
Significant changes in the $E$-vector position angles of 
90$^{\circ}$--105$^{\circ}$ are observed from the partial ``ring-like structure'' 
particularly at the NE and SW components to the ``centrally concentrated component''.  
A high polarization percentage of ${\gtrsim}10\%$ is derived 
in the central 200 AU both in the ``ring-like structure'' and the 
``centrally concentrated component''. 

\item	We have analyzed origin 
of the polarized emission and three main mechanisms 
to align dust grains and/or to produce polarization: 
(i) magnetic field, (ii) self-scattering, 
and (iii) anisotropic radiation. 
Based on a comparison with the available data, 
the magnetic field scenario 
appears to explain best the observed characteristics. 
The $E$-vector orientations in 
the ``centrally concentrated component'' are 
also consistent with the self-scattering model 
with an optically thick disk-like structure. 
However, the polarization degree appears to be significantly higher 
than the expected value for the self-scattering mechanism.

\item	The ``arm-like structure'' is also observed 
on large scales ($\approx$2000 AU). 
The detected polarization $E$-vectors are aligned 
along the minor axis of the ``arm-like structure'', 
which is consistent with magnetically aligned dust 
in the field parallel to the major axis of the arm.

\item	We also performed a MHD simulation to further explore 
the magnetic field structures at the vicinity of the protostar. 
After taking into account the viewing angle effect 
and the misalignment between 
the initial magnetic field orientation 
and core rotational axis (i.e., outflow axis adjusted for the case of MMS 6), 
the calculated projected polarization image reproduces 
most of the characteristic structures observed in the ALMA images 
including the ``ring-like structure'' and the ``arm-like structure''. 
This agreement further supports 
that the origin of the dust alignment for 
this source can be explained by a toroidal wrapping of the magnetic fields. 
However, the observed ``centrally concentrated component'' 
is not reproduced due to the limitations 
of the current MHD simulations. 

\end{enumerate}

This is the first time that study of the polarization 
could be done in such details in a protostellar core 
and that new question and issues have come up that 
need to be further scrutinized. From the observational side, 
multi-wavelength polarization experiments with similarly high angular resolution  
will be crucial in order to assess the hybrid models 
by investigating contributions from the self-scattering 
and dust alignment by anisotropic radiation. 
From the theoretical side, taking into account the optical depth effect 
and the contribution from the central protostar will help to further study  
the polarization properties in the central 200 AU. 
These improvements will be done in our next papers.

\acknowledgments
\noindent

{\it Acknowledgements.}
We thank the anonymous referee for providing very helpful 
comments and suggestions. S. Takahashi thanks Zhi-Yun Li, 
A. Kataoka, P. Cox, S. Koga, Y.-N., Su, G.H.-M. Bertrang, 
K. Saigo, and S. Iguchi for fruitful comments and discussion. 
We also thank T. Michiyama and B. Dent for providing us information 
regarding polarization calibration accuracy and observing condition, respectively.  
This paper makes use of the following ALMA data: ADS/JAO.ALMA \#2015.1.00341.S. 
ALMA is a partnership of ESO (representing its member states), NSF (USA) 
and NINS (Japan), together with NRC (Canada), MOST and ASIAA (Taiwan), 
and KASI (Republic of Korea), in cooperation with the Republic of Chile. 
The Joint ALMA Observatory is operated by ESO, AUI/NRAO and NAOJ. 
This work was supported by JSPS KAKENHI Grant Numbers 17K05387 and 15K05032. 
This research used computational resources from the high-performance
computing infrastructure (HPCI) system provided by the Cyberscience 
Center, Tohoku University, and the Cybermedia Center, Osaka
University through the HPCI System Research Project 
(Project ID: hp170047, hp180001). 
P.T.P. Ho acknowledges support from MOST 105-2112-M001-025-MY3. 

{\it Facilities:} \facility{ALMA}.

\clearpage

\begin{figure}
\epsscale{1.0}
\plotone{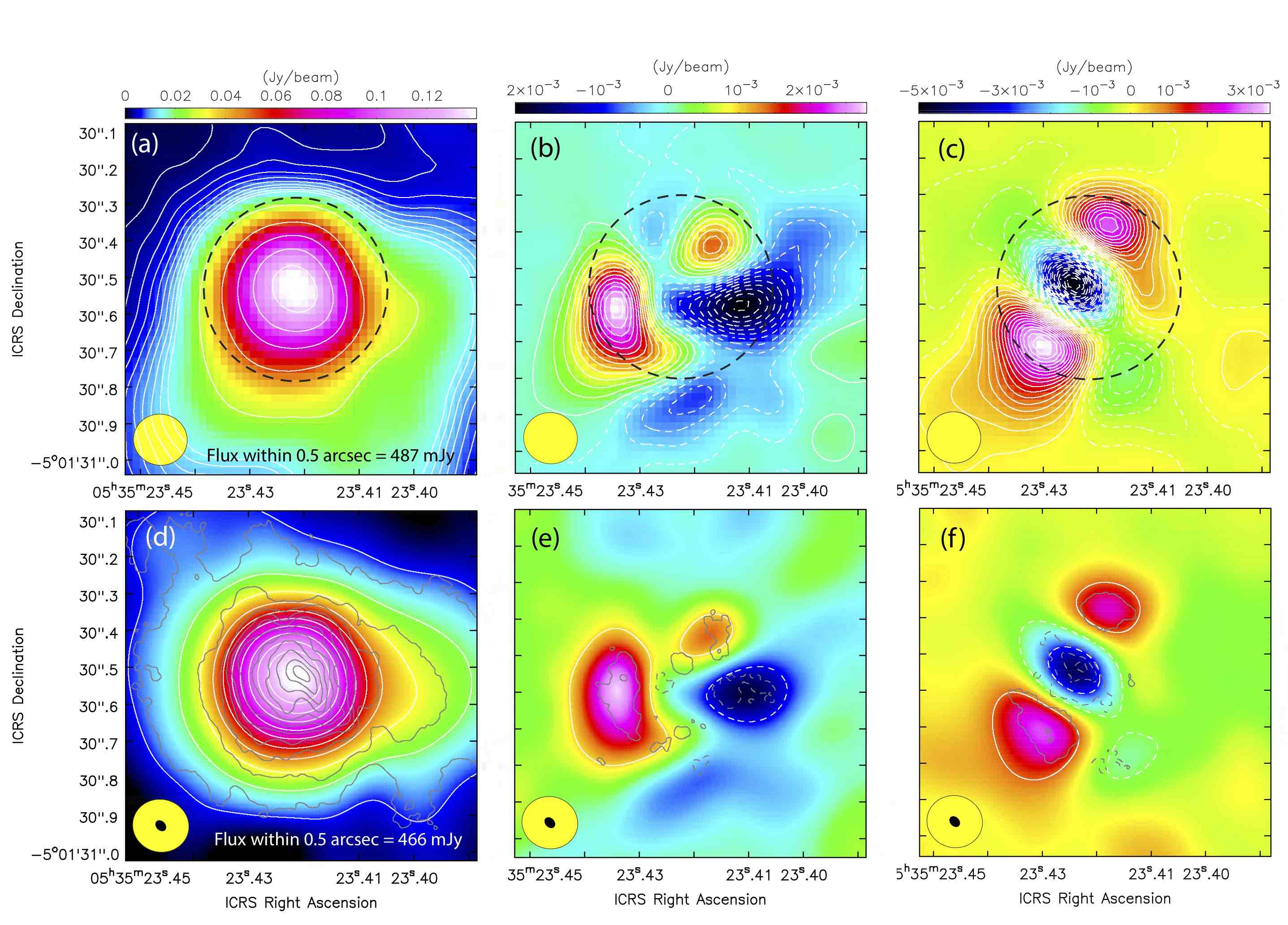}
\caption{\scriptsize Stokes $I$, $Q$, and $U$ images obtained from the low angular 
resolution data sets (a, b, and c) and the high angular 
resolution data sets (d, e, and f), respectively. 
{\bf Low angular resolution images;(a), (b), and (c)}: 
Both color scale and white contours show the Stokes $I$, $Q$, and $U$ intensities. 
Contour levels of Stokes $I$ correspond to 5$\sigma$, 15$\sigma$, 25$\sigma$, 
35$\sigma$, 45$\sigma$, 55$\sigma$, 65$\sigma$, 75$\sigma$, 85$\sigma$, 95$\sigma$, 
105$\sigma$, 200$\sigma$, 400$\sigma$, 600$\sigma$, 800$\sigma$, and 1000$\sigma$ 
(1$\sigma$=130 $\mu$Jy beam$^{-1}$). 
Contour level of Stokes $Q$ starts from $-105{\sigma}$ up to 135$\sigma$  
with the interval of 10$\sigma$ (1$\sigma$=20 $\mu$Jy beam$^{-1}$). 
Contour level of Stokes $U$ starts from $-265{\sigma}$ up to 165$\sigma$ 
with the interval of 10$\sigma$ (1$\sigma$=20 $\mu$Jy beam$^{-1}$).
{\bf High angular resolution images;(d), (e), and (f)}: Gray contours show 
the original Stokes $I$, $Q$, and $U$ intensities obtained 
from the high angular resolution data set, while both white contours and color 
show the high angular resolution Stokes $I$, $Q$, and $U$ convolved with 
the synthesized beam size of the low angular resolution images 
of $0''.15{\times}0''.14$ (P.A.=80$^{\circ}$). 
Contour level of Stokes $I$ starts from 5$\sigma$ up to 105$\sigma$ 
with the interval of 10$\sigma$ (1$\sigma$=63 $\mu$Jy beam$^{-1}$ 
for the gray contours and 1$\sigma$=1.3 mJy beam$^{-1}$ for the white contours). 
Contour levels of Stokes $Q$ correspond to $-5{\sigma}$, 5$\sigma$, and 15$\sigma$ 
(1$\sigma$=21 $\mu$Jy beam$^{-1}$ for the gray contours 
and 1$\sigma$=200 $\mu$Jy beam$^{-1}$ for the white contours). 
Contour levels of Stokes $U$ correspond to $-25{\sigma}$, $-15{\sigma}$, 
$-5{\sigma}$, 5$\sigma$, and 15$\sigma$ 
(1$\sigma$=21 $\mu$Jy beam$^{-1}$ for the gray contours 
and 1$\sigma$=200 $\mu$Jy beam$^{-1}$ for the white contours). 
Stokes $I$ fluxes measured within the central $0''.5$ 
i.e., dashed open circles in Figure (a), (b), and (c) 
are written in the bottom right corners in Figures (a) and (d). 
The synthesized beam sizes for the high and low angular resolution 
images are denoted in the bottom left corners with black 
and yellow filled ellipses, respectively.} 
\end{figure}

\clearpage

\begin{figure}
\epsscale{0.8}
\plotone{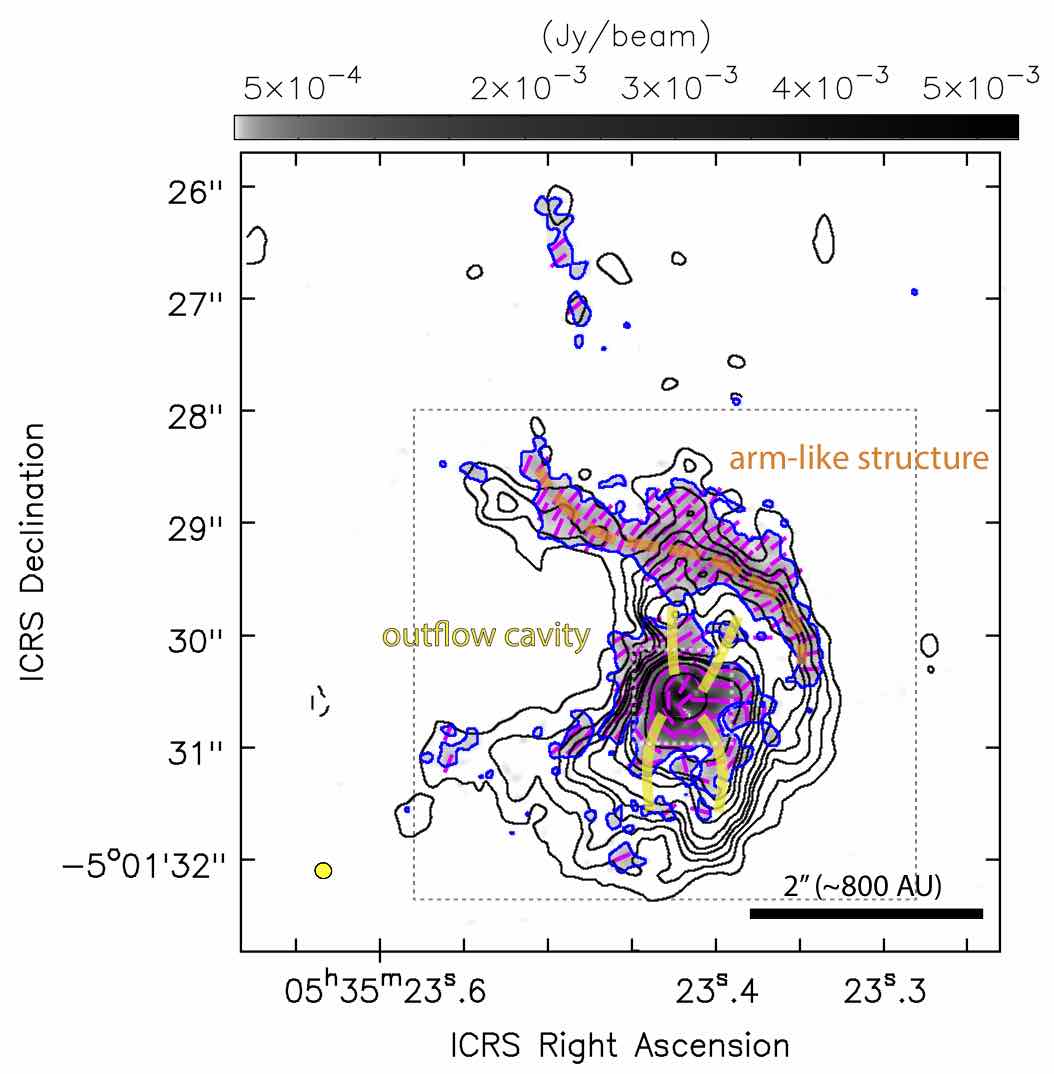}
\caption{The full polarization 1.1 mm continuum images obtained from 
the low angular resolution data set. Gray scale presents the linearly 
polarized flux overlaid with the Stokes $I$ (black contours) and $E$-vectors (magenta lines). 
The 5$\sigma$ level of the polarized flux are denoted in the blue lines 
(1$\sigma$=28 $\mu$Jy beam$^{-1}$). 
``Arm-like structure'' and ``outflow cavity'' 
explained in the main text are 
denoted with the orange and yellow guide lines. 
The large and small squares in the image correspond 
to the image sizes of Figures 3 and 4(a), respectively. 
The contour levels of the Stokes $I$ 
correspond to $-3{\sigma}$, 3$\sigma$, 6$\sigma$, 
9$\sigma$, 12$\sigma$, 
15$\sigma$, 20$\sigma$, 30$\sigma$, 40$\sigma$, 
50$\sigma$, 100$\sigma$, 200$\sigma$, 
and 400$\sigma$ (1$\sigma$=130 $\mu$Jy beam$^{-1}$). 
The synthesized beam size is denoted in 
the bottom-left corner with a filled yellow ellipse.}
\end{figure}

\clearpage

\begin{figure}
\epsscale{1.00}
\plotone{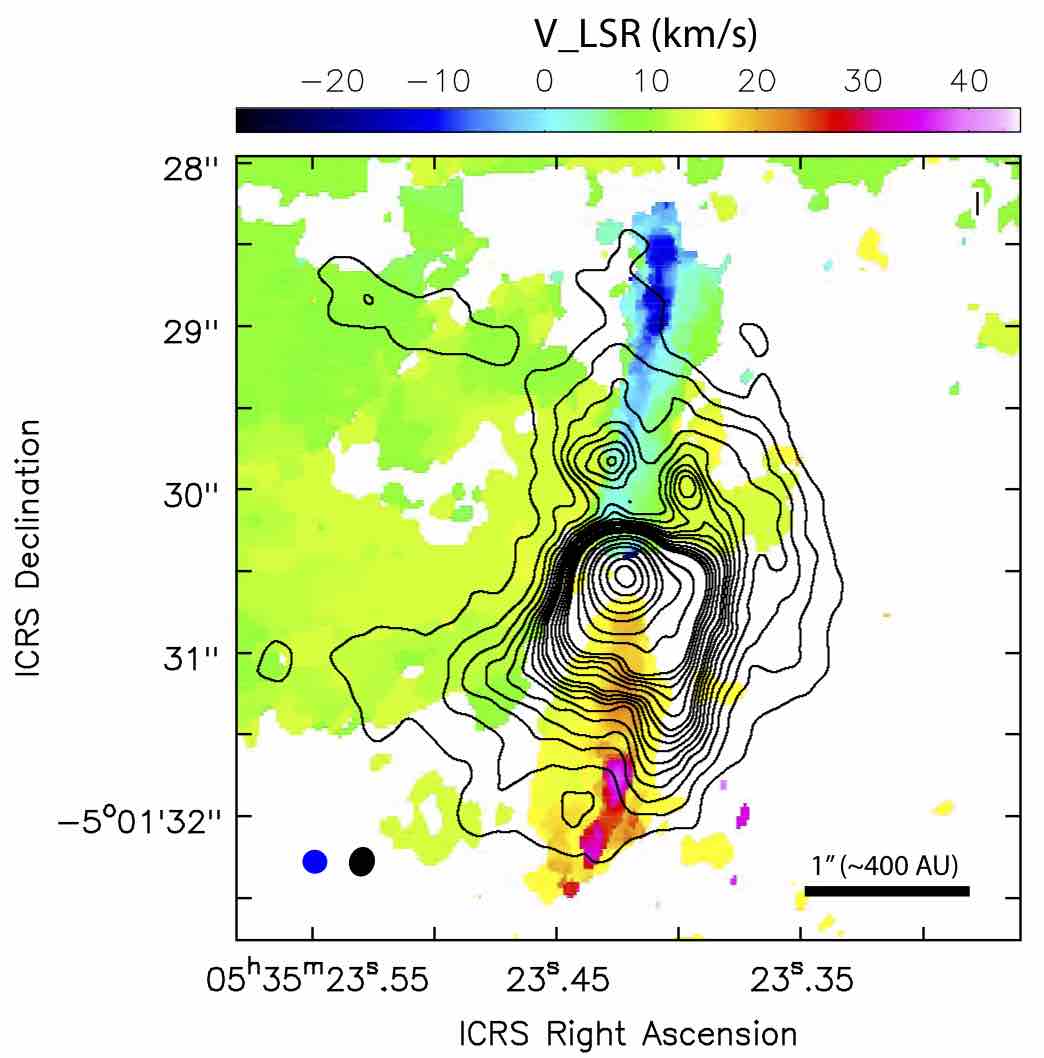}
\caption{The mean velocity map calculated from the CO $J$=2--1 data (color) 
overlaid with the 1.1 mm dust continuum emission obtained 
from the ALMA low angular resolution data (contours) 
for the central region shown in Figure 2. 
The contour levels start at $-5{\sigma}$ with an interval of 5$\sigma$ up to 50$\sigma$, 
then continues with 60$\sigma$, 70$\sigma$, 80$\sigma$, 90$\sigma$, 100$\sigma$, 
150$\sigma$, 200$\sigma$, 300$\sigma$, 400$\sigma$, 600$\sigma$, 
800$\sigma$, and 1000$\sigma$ (1$\sigma$=130 $\mu$Jy beam$^{-1}$). 
The beam sizes of the 1.1 mm Stokes $I$ and CO (2--1) emission are 
presented in the bottom left corner with a filled blue and black ellipses, 
respectively.}
\label{}
\end{figure}

\clearpage

\begin{landscape}
\begin{figure}
\epsscale{1.0}
\plotone{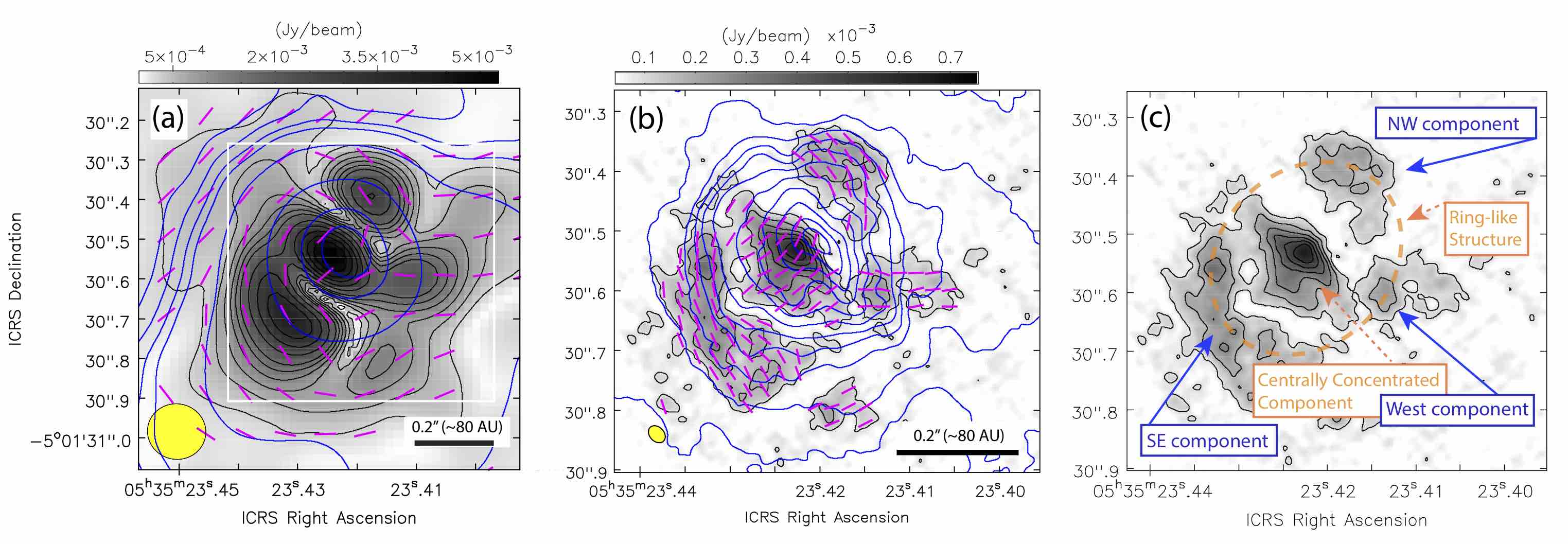}
\caption{\small Full polarization 1.1 mm continuum images at the central 400 AU 
obtained from the low angular resolution (Figure 4a) 
and high angular resolution (Figure 4b) data sets. 
Gray scale and black contours present the linearly polarized flux, 
blue contours and magenta lines present the Stokes $I$ emission 
and the $E$-vectors. 
(a) Full polarization image obtained from 
the inner squared region of Figure 2. 
Contour levels of the Stokes $I$ and polarized emission correspond 
to 6$\sigma$, 9$\sigma$, 12$\sigma$, 15$\sigma$, 20$\sigma$, 30$\sigma$, 
40$\sigma$, 50$\sigma$, 100$\sigma$, 200$\sigma$, 400$\sigma$, 
and 800$\sigma$ (1$\sigma$=130 $\mu$Jy beam$^{-1}$), 
and 10$\sigma$, 20$\sigma$, 
30$\sigma$, 40$\sigma$, 50$\sigma$, 60$\sigma$, 
80$\sigma$, 100$\sigma$, 120$\sigma$, 140$\sigma$, 
160$\sigma$, and 180$\sigma$ 
(1$\sigma$=28$\mu$Jy beam$^{-1}$), respectively. 
(b) Full polarization image obtained from 
the high angular resolution image. Zoomed 
image within the white squared region in Figure 4(a). 
Contour levels of the Stokes $I$ 
and polarized emission start at 5$\sigma$ 
and 3$\sigma$ with intervals of 10$\sigma$ 
and 3$\sigma$, respectively (1$\sigma$=63 $\mu$Jy beam$^{-1}$ 
for the Stokes $I$ emission and 1$\sigma$=30 $\mu$Jy beam$^{-1}$ 
for the $PI$). 
(c) High angular resolution polarized 
intensity image, same as Figure 4(b), overlaid with 
structure information referred in the main text. 
For all the images, the synthesized beam sizes 
are denoted in the bottom left corners 
of Figures (a) and (b) with filled yellow ellipses.} 
\end{figure}
\end{landscape}

\begin{landscape}
\begin{figure}
\epsscale{1.0}
\plotone{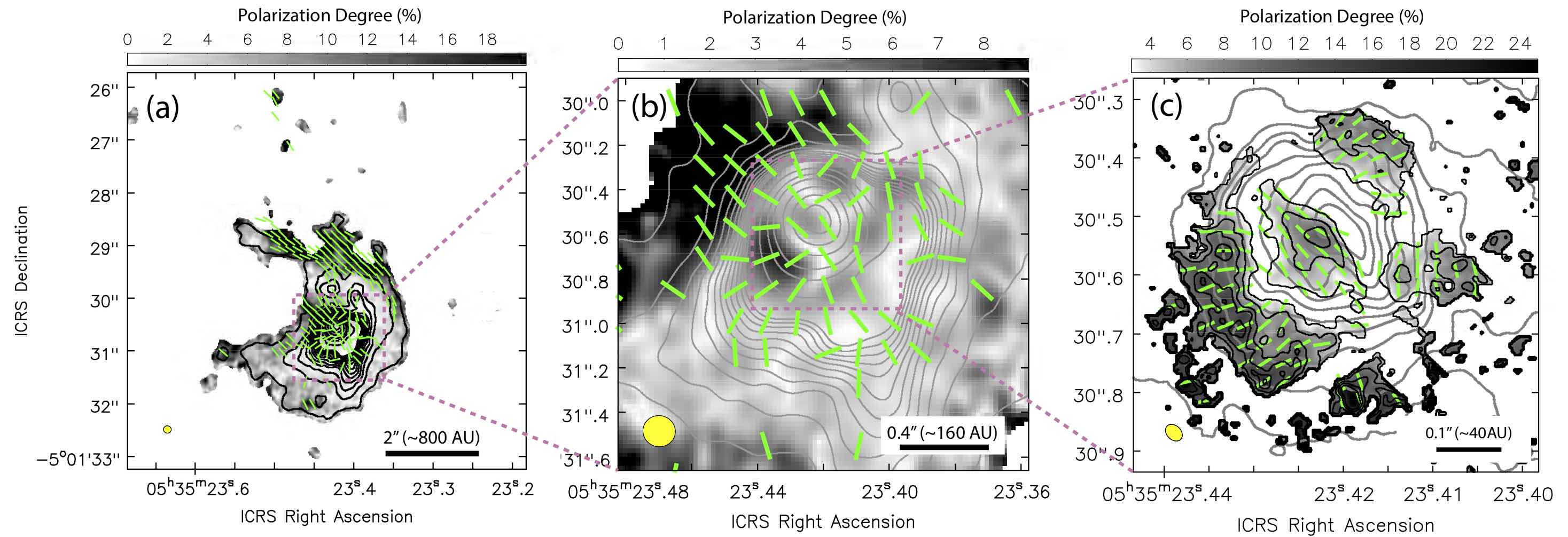}
\caption{Polarization degree maps (gray scale) overlaid with 
the Stokes $I$ emission (gray contours) and $B$-vectors (green vectors). 
(a) Large-scale image made from the low angular resolution data set. 
Contour levels of the Stokes $I$ correspond to 5$\sigma$, 15$\sigma$, 
25$\sigma$, 35$\sigma$, 45$\sigma$, 55$\sigma$, 65$\sigma$, 75$\sigma$, 85$\sigma$, 95$\sigma$, 105$\sigma$, 115$\sigma$, 200$\sigma$, 300$\sigma$, 400$\sigma$, 600$\sigma$, 800$\sigma$ and 1000$\sigma$ (1$\sigma$=130 $\mu$Jy beam$^{-1}$). (b) Zoomed image of panel (a). (c) Zoomed image of panel (b) obtained from the high angular resolution data set. The contour levels of the Stokes $I$ (gray) and the polarization degree (black) start from 5$\sigma$ with 10$\sigma$ interval (1$\sigma$=63 $\mu$Jy beam$^{-1}$) and start from 3\% with the interval of 3\%, respectively. 
The synthesized beam sizes are denoted in the bottom left corners 
with filled yellow ellipses.} 
\end{figure}
\end{landscape}

\clearpage

\begin{figure}
\epsscale{1.00}
\plotone{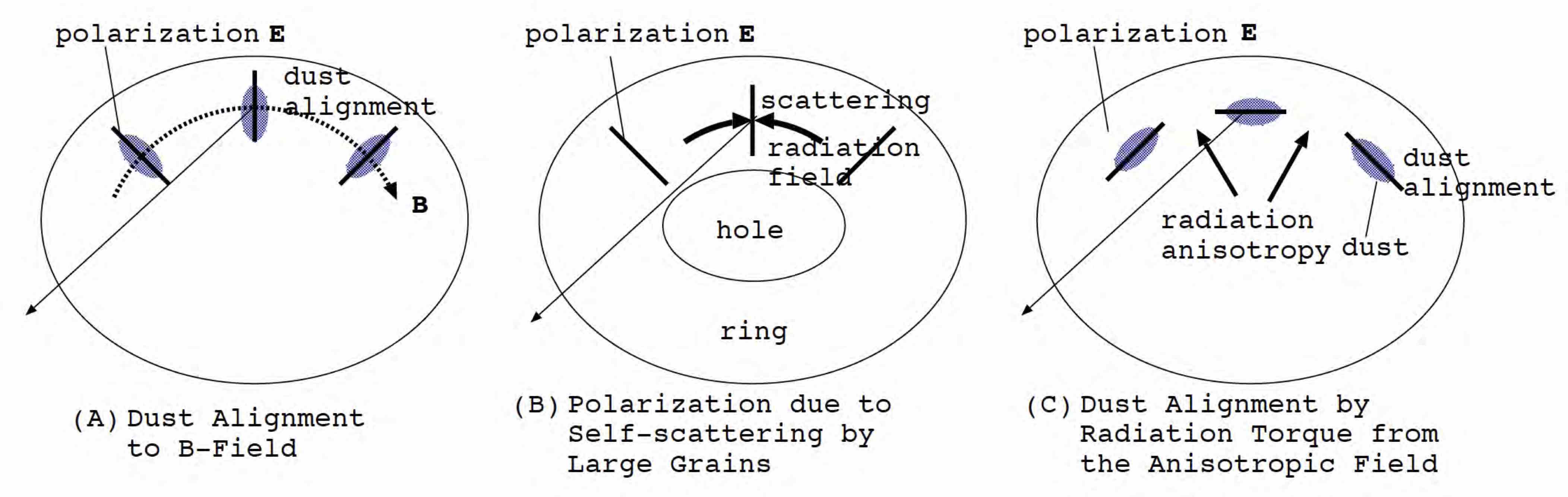}
\caption{Schematic pictures to explain the dust grain alignment 
and the direction of the E-vectors: 
(A) Dust alignment caused by the magnetic alignment, 
(B) Self-scattering model, and (C) Dust alignment 
by the anisotropic radiation.}
\label{}
\end{figure}

\begin{figure}
\epsscale{1.0}
\plotone{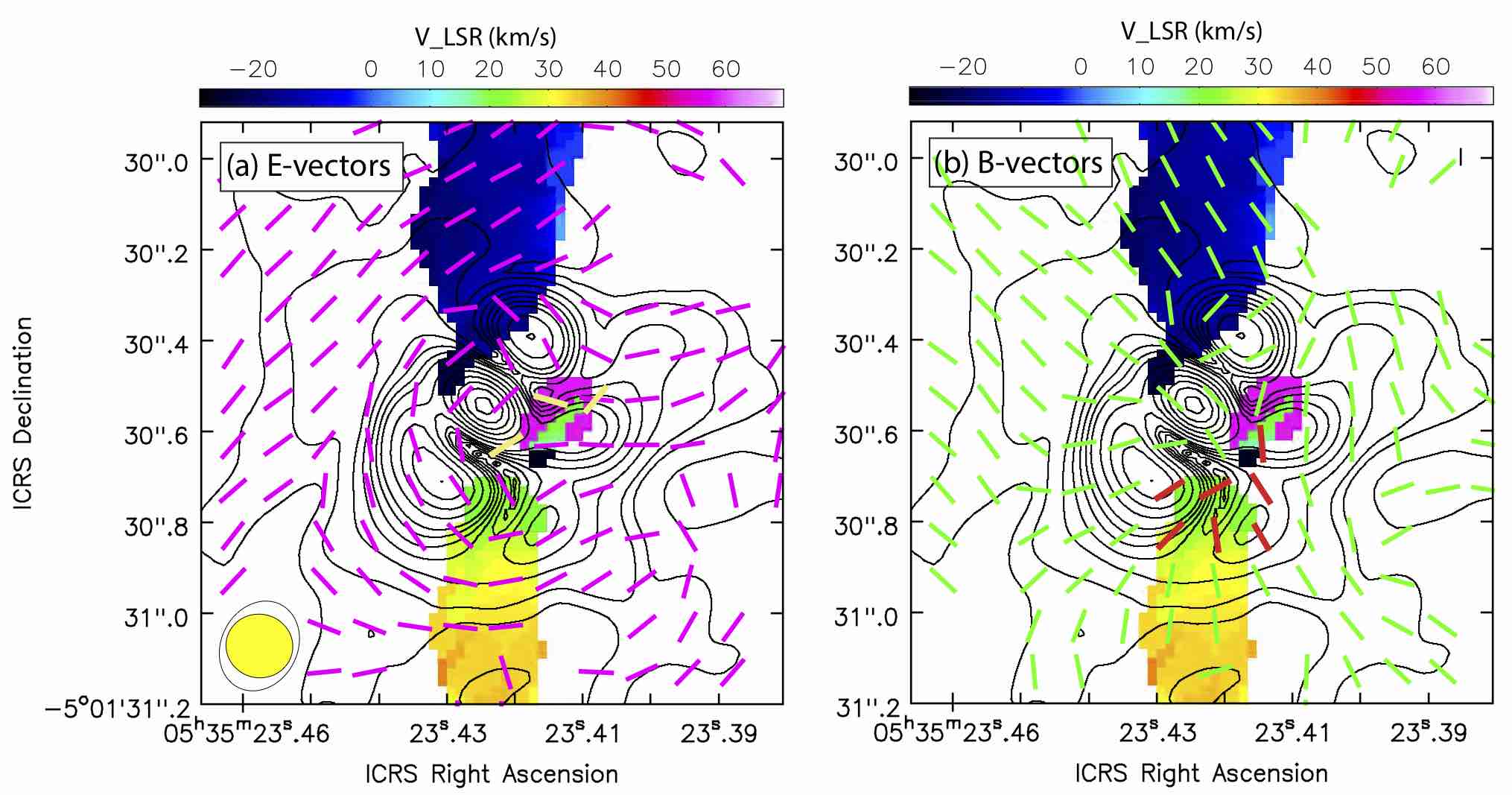}
\caption{The mean velocity map calculated from the SiO (5--4) observations (color) 
overlaid with the $PI$ (contours) obtained from 
the low angular resolution image, and (a) $E$-vectors and (b) $B$-vectors. 
The contour levels correspond to 5$\sigma$, 10$\sigma$, 15$\sigma$, 
30$\sigma$, 40$\sigma$, 50$\sigma$, 60$\sigma$, 80$\sigma$, 
100$\sigma$, 120$\sigma$, 140$\sigma$, 160$\sigma$, and 180$\sigma$ 
(1$\sigma$=28 $\mu$Jy beam$^{-1}$). 
The synthesized beam sizes of the SiO (5--4) and 1.1 mm images
are denoted in open black and yellow filled ellipses, respectively.}
\label{}
\end{figure}

\begin{figure}
\epsscale{0.8}
\plotone{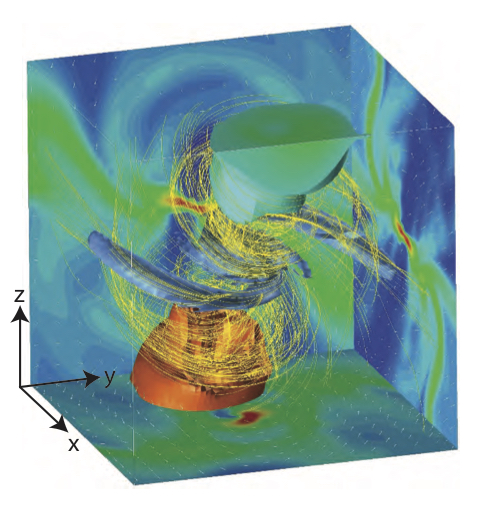}
\caption{Three-dimensional view of structures of magnetic field lines (yellow), 
gas density (blue) and outflow (red and blue). 
The density and velocity distributions 
on the $x$=0, $y$=0 and $z$=0 cutting plane are projected onto 
each wall surface. The box scale is 200 AU. 
For visualization purpose, we assumed a line of sight 
in the direction of ${\theta}=60^{\circ}$ and 
${\phi}=15^{\circ}$ in the ordinary three-dimensional spherical coordinate. 
Here, ${\theta}=0^{\circ}$ coincides with the $z$-axis or 
the direction of initial magnetic field.}
\label{}
\end{figure}

\begin{figure}
\epsscale{1.0}
\plotone{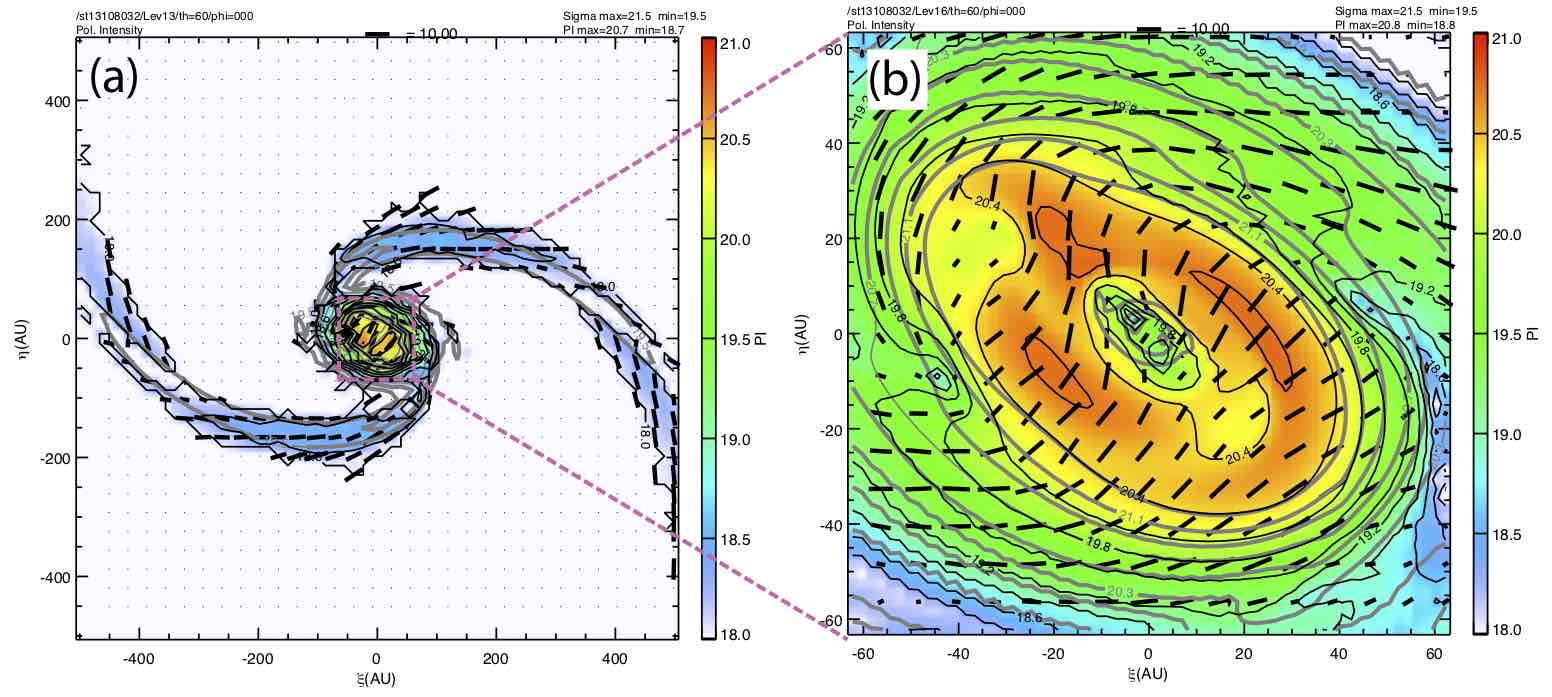}
\caption{Projected polarization images produced 
by the MHD simulation presented in Figure 8. 
Viewing angle of $\theta$=60$^{\circ}$ and $\phi$=0$^{\circ}$ are 
used to integrate over the system. 
Panel (a) shows the large scale image (central 900 AU region), 
while (b) shows a zoomed image (central 120 AU). 
Definitions of the coordinate and image size follows 
those from \cite{2011PASJ...63..147T}. 
Column density, magnetic field orientations, 
and $PI$ are denoted in gray contours, 
black lines, color images overlaid with black contours, respectively. 
Polarization (color) is calculated from the relative Stokes 
parameters \citep{1985ApJ...290..211L,2011PASJ...63..147T}, 
assuming that the dust properties, dust alignment degree, 
and dust temperature are all uniform. 
Thus, the unit of the $PI$ is arbitrary 
and the color indicates only the relative difference of the $PI$.}
\label{}
\end{figure}

\begin{deluxetable}{lccc}
\tabletypesize{\scriptsize}
\tablecaption{ALMA Observing Parameters for the Full Polarization Experiments}
\tablewidth{0pt}
\tablehead{
\colhead{Parameters} & \colhead{X1aa2 \& X269f} & \colhead{X368f, X435c, \& X4b9d} & \colhead{X336 \& X3d1d}}
\startdata
Observing date (YYYY-MM-DD)				& 2015-10-29	& 2016-10-09		& 2016-10-11		\\
Number of antennas						& 38 			& 42				& 44 				\\
Primary beam size (arcsec)      		& 21			& 21			& 21				\\
PWV (mm)								& 1.3 --1.9 	& 0.44 -- 0.90	    & 0.44 -- 0.57		\\
Phase stability rms (degree)\tablenotemark{a} 			& 16--23		& 	16 -- 56		& 10 --15	\\
Polarization calibrator					& J0522-3627	& J0522-3627	& J0522-3627	 	\\
Bandpass calibrators  					& J0423-0120 	& J0510+1800		& J0510+1800	 	\\
Flux calibrator  						& J0423-0120	& J0423-0120	& J0423-0120		\\
Phase calibrators (separation from the target)\tablenotemark{b}  & J0541-0541 (1.7$^{\circ}$)	& J0532-0307 (2.1$^{\circ}$) & J0541-0541 (1.7$^{\circ}$)	\\
Central frequency USB/LSB (GHz)    & 	257 / 273 & 257 / 273 & 257 / 273	\\
Total continuum bandwidth; USB+LSB (GHz)	& 7.5		& 7.5 			& 7.5				\\
Projected baseline ranges (k$\lambda$)  & 76 --14700	& 16 -- 3200		& 16 -- 3200		\\
Maximum recoverable size (arcsec)\tablenotemark{c}		& 0.22				& 1.0 				& 1.0	\\
On-source time (minutes)      			& 41			& 12.5				& 13				\\       
Synthesized beam size (arcsec)\tablenotemark{d}			& $0''.03{\times}0''.02$ (P.A.=43$^{\circ}$)				& \multicolumn{2}{c}{$0''.15{\times}0''.14$ (P.A.=80$^{\circ}$)}\\
RMS noise level of Stokes $I$\tablenotemark{e}, $Q$, and $U$ ($\mu$Jy beam$^{-1}$)	&		63, 21, and 21		& 
\multicolumn{2}{c}{130, 20, and 20}	\\
\enddata
\tablenotetext{a}{Antenna-based phase differences measured on the bandpass calibrator.}
\tablenotetext{b}{The phase calibrator was observed every 1 min. and 8 min. for the high angular and low angular resolution observations, respectively.}
\tablenotetext{c}{Our observations were insensitive to emission more extended than this size scale structure at the 10\% level (Wilner \& Welch 1994).}
\tablenotetext{d}{The natural weighting and the Briggs weighting (robust parameter of 0.5) are used for the high and low angular resolution imaging, respectively.}
\tablenotetext{e}{Stokes $I$ images are dynamic range limited.}
\end{deluxetable}

\begin{deluxetable}{lccccc}
\tabletypesize{\scriptsize}
\tablecaption{Multiple Gaussian Fitting Results}
\tablewidth{0pt}
\tablehead{
\colhead{} & \colhead{R.A.} & \colhead{Decl.} & \colhead{Deconvolved Size, P.A.}	& \colhead{Peak Intensity}	& \colhead{Flux Density} \\
\colhead{} & \colhead{(J2000)} & \colhead{(J2000)} & \colhead{(milli-arcsec, deg)}	& \colhead{(mJy beam$^{-1}$)}	& \colhead{(mJy)}}
\startdata
(i) Extended component	& 05 35 23.4220	& -05 01 30.532	& $321{\pm}4{\times}312{\pm}4$, 134${\pm}18$	& 5.5${\pm}$0.1	& 724${\pm}$8 	\\
(ii) Elongated component & 05 35 23.4182  & -05 01 30.553 & $314{\pm}37{\times}55{\pm}6$, 39${\pm}$2	& 1.1${\pm}$0.1 & 28${\pm}$3  	\\
(iii) Compact	component & 05 35 23.4201 & -05 01 30.515 & $54{\pm}32{\times}13{\pm}15$, $72{\pm}46$	& 0.7${\pm}$0.2 & 1.5${\pm}$0.8	\\
\enddata
\end{deluxetable}

\begin{deluxetable}{lcccc}
\tabletypesize{\scriptsize}
\tablecaption{Physical Properties of Fitted Components}
\tablewidth{0pt}
\tablehead{
\colhead{Component} & \colhead{Source Size} & \colhead{$M_{\rm{H_2}}$} & \colhead{$n_{\rm{H_2}}$}	& \colhead{$n_{\rm{H_2}}$} \\
\colhead{} & \colhead{(AU)} & \colhead{($M_{\odot}$)} & \colhead{(cm$^{-2}$)}	& \colhead{(cm$^{-3}$)}}	
\startdata
(i) Extended component		& $133({\pm}1){\times}129({\pm}1)$	 & 9.5e-01 (${\pm}$1.0e-02)  & 6.4e+25 (${\pm}$6.9e+23) & 4.3e+10 (${\pm}$4.7e+08)    \\
(ii) Elongated component	& $130({\pm}15){\times}23({\pm}3)$   & 3.3e-03 (${\pm}$3.7e-04)  & 4.4e+24 (${\pm}$4.9e+23) & 4.2e+09 (${\pm}$4.7e+08)    \\
(iii) Compact component 	& $22({\pm}13){\times}5({\pm}6)$	 & 1.3e-04 (${\pm}$6.6e-05)  & 5.6e+24 (${\pm}$2.9e+24) & 3.1e+10 (${\pm}$1.6e+10)    \\
\enddata
\end{deluxetable}

\begin{deluxetable}{lccc}
\tabletypesize{\scriptsize}
\tablecaption{Comparisons of Three Proposed Polarization Mechanism at the Central 200 AU\tablenotemark{a}}
\tablewidth{0pt}
\tablehead{
\colhead{} & \colhead{Magnetic Field} & \colhead{Self-scattering} & \colhead{Anisotropic radiation}
}
\startdata
(i) Spatial distribution difference between Stokes $I$ and $PI$ & ${\bigcirc}$ & {\large {$\vartriangle$}} & {\large ${\times}$} \\
(ii) Peak positional offset between the Stoke $I$ and $PI$		& {\large $\vartriangle$}	 & ${\bigcirc}$  & {\large $\vartriangle$} \\
(iii) ring-like depolarized region (ring-like gap)			& ${\bigcirc}$	 & {\large $\vartriangle$}	& {\large ${\times}$}  \\
(iv) Change of the $E$-vector orientations across the ring-like gap & ${\bigcirc}$ & ${\bigcirc}$ & {\large $\vartriangle$}  \\
(v) High polarization percentage of $\gtrsim$10\%			& {\large $\vartriangle$}	& {\large ${\times}$} 	& {\large $\vartriangle$}  \\
\enddata
\tablenotetext{a}{$\bigcirc$: Result can be reproduced by the existing models, 
{\large $\vartriangle$}: Result could be explained by the proposed model qualitatively, 
but no clear cases are reported as publication before. 
{\large $\times$}: Result cannot be explained by the existing models.}
\end{deluxetable}


\begin{thebibliography}{}
\bibitem[Alves et al.(2018)]{2018A&A...616A..56A} Alves, F.~O., Girart, J.~M., Padovani, M., et al.\ 2018, \aap, 616, A56 
\bibitem[Andr{\'e} et al.(2010)]{2010A&A...518L.102A} Andr{\'e}, P., Men'shchikov, A., Bontemps, S., et al.\ 2010, \aap, 518, L102 
\bibitem[Anglada et al.(1998)]{1998AJ....116.2953A} Anglada, G., Villuendas, E., Estalella, R., et al.\ 1998, \aj, 116, 2953 
\bibitem[Arce et al.(2007)]{2007prpl.conf..245A} Arce, H.~G., Shepherd, D., Gueth, F., et al.\ 2007, Protostars and Planets V, 245 
\bibitem[Bacciotti et al.(2018)]{2018ApJ...865L..12B} Bacciotti, F., Girart, J.~M., Padovani, M., et al.\ 2018, \apjl, 865, L12 
\bibitem[Banerjee \& Pudritz(2006)]{2006ApJ...641..949B} Banerjee, R., \& Pudritz, R.~E.\ 2006, \apj, 641, 949 
\bibitem[Chini et al.(1997)]{1997ApJ...474L.135C} Chini, R., Reipurth, B., Ward-Thompson, D., et al.\ 1997, \apjl, 474, L135 
\bibitem[Commer{\c c}on et al.(2010)]{2010A&A...510L...3C} Commer{\c c}on, B., Hennebelle, P., Audit, E., Chabrier, G., \& Teyssier, R.\ 2010, \aap, 510, L3 
\bibitem[Cortes et al.(2016)]{2016ApJ...825L..15C} Cortes, P.~C., Girart, J.~M., Hull, C.~L.~H., et al.\ 2016, \apjl, 825, L15 
\bibitem[Cox et al.(2018)]{2018ApJ...855...92C} Cox, E.~G., Harris, R.~J., Looney, L.~W., et al.\ 2018, \apj, 855, 92 
\bibitem[Crutcher(2012)]{2012ARA&A..50...29C} Crutcher, R.~M.\ 2012, \araa, 50, 29 
\bibitem[Davis \& Greenstein(1951)]{1951ApJ...114..206D} Davis, L., Jr., \& Greenstein, J.~L.\ 1951, \apj, 114, 206 
\bibitem[Dent et al.(2019)]{2019MNRAS.482L..29D} Dent, W.~R.~F., Pinte, C., Cortes, P.~C., et al.\ 2019, \mnras, 482, L29 
\bibitem[Galli \& Shu(1993)]{1993ApJ...417..243G} Galli, D., \& Shu, F.~H.\ 1993, \apj, 417, 243 
\bibitem[Girart et al.(2006)]{2006Sci...313..812G} Girart, J.~M., Rao, R., \& Marrone, D.~P.\ 2006, Science, 313, 812 
\bibitem[Girart et al.(2009)]{2009Sci...324.1408G} Girart, J.~M., Beltr{\'a}n, M.~T., Zhang, Q., Rao, R., \& Estalella, R.\ 2009, Science, 324, 1408 
\bibitem[Girart et al.(2013)]{2013ApJ...772...69G} Girart, J.~M., Frau, P., Zhang, Q., et al.\ 2013, \apj, 772, 69 
\bibitem[Girart et al.(2018)]{2018ApJ...856L..27G} Girart, J.~M., Fern{\'a}ndez-L{\'o}pez, M., Li, Z.-Y., et al.\ 2018, \apjl, 856, L27 
\bibitem[Gold(1952)]{1952MNRAS.112..215G} Gold, T.\ 1952, \mnras, 112, 215 
\bibitem[Harris et al.(2018)]{2018ApJ...861...91H} Harris, R.~J., Cox, E.~G., Looney, L.~W., et al.\ 2018, \apj, 861, 91 
\bibitem[Hildebrand(1988)]{1988QJRAS..29..327H} Hildebrand, R.~H.\ 1988, \qjras, 29, 327 
\bibitem[Ho et al.(2004)]{2004ApJ...616L...1H} Ho, P.~T.~P., Moran, J.~M., \& Lo, K.~Y.\ 2004, \apjl, 616, L1 
\bibitem[Hull et al.(2014)]{2014ApJS..213...13H} Hull, C.~L.~H., Plambeck, R.~L., Kwon, W., et al.\ 2014, \apjs, 213, 13 
\bibitem[Hull et al.(2017)]{2017ApJ...847...92H} Hull, C.~L.~H., Girart, J.~M., Tychoniec, {\L}., et al.\ 2017, \apj, 847, 92 
\bibitem[Hull et al.(2018)]{2018ApJ...860...82H} Hull, C.~L.~H., Yang, H., Li, Z.-Y., et al.\ 2018, \apj, 860, 82 
\bibitem[Kataoka et al.(2012)]{2012ApJ...761...40K} Kataoka, A., Machida, M.~N., \& Tomisaka, K.\ 2012, \apj, 761, 40 
\bibitem[Kataoka et al.(2015)]{2015ApJ...809...78K} Kataoka, A., Muto, T., Momose, M., et al.\ 2015, \apj, 809, 78 
\bibitem[Kataoka et al.(2016)]{2016ApJ...831L..12K} Kataoka, A., Tsukagoshi, T., Momose, M., et al.\ 2016, \apjl, 831, L12 
\bibitem[Kataoka et al.(2017)]{2017ApJ...844L...5K} Kataoka, A., Tsukagoshi, T., Pohl, A., et al.\ 2017, \apjl, 844, L5 
\bibitem[Keene et al.(1982)]{1982ApJ...252L..11K} Keene, J., Hildebrand, R.~H., \& Whitcomb, S.~E.\ 1982, \apjl, 252, L11 
\bibitem[Koch et al.(2018)]{2018ApJ...855...39K} Koch, P.~M., Tang, Y.-W., Ho, P.~T.~P., et al.\ 2018, \apj, 855, 39 
\bibitem[Kounkel et al.(2017)]{2017ApJ...834..142K} Kounkel, M., Hartmann, L., Loinard, L., et al.\ 2017, \apj, 834, 142 
\bibitem[Kwon et al.(2018)]{2018arXiv180507348K} Kwon, W., Stephens, I., Tobin, J., et al.\ 2018, arXiv:1805.07348 
\bibitem[Lai et al.(2003)]{2003ApJ...598..392L} Lai, S.-P., Girart, J.~M., \& Crutcher, R.~M.\ 2003, \apj, 598, 392 
\bibitem[Lazarian(2007)]{2007JQSRT.106..225L} Lazarian, A.\ 2007, \jqsrt, 106, 225 
\bibitem[Lazarian \& Hoang(2007)]{2007ApJ...669L..77L} Lazarian, A., \& Hoang, T.\ 2007, \apjl, 669, L77 
\bibitem[Lee \& Draine(1985)]{1985ApJ...290..211L} Lee, H.~M., \& Draine, B.~T.\ 1985, \apj, 290, 211 
\bibitem[Lee et al.(2018)]{2018ApJ...854...56L} Lee, C.-F., Li, Z.-Y., Ching, T.-C., Lai, S.-P., \& Yang, H.\ 2018, \apj, 854, 56 
\bibitem[Li et al.(2014)]{2014ApJ...793..130L} Li, Z.-Y., Krasnopolsky, R., Shang, H., \& Zhao, B.\ 2014, \apj, 793, 130 
\bibitem[Machida et al.(2005)]{2005MNRAS.362..369M} Machida, M.~N., Matsumoto, T., Tomisaka, K., \& Hanawa, T.\ 2005, \mnras, 362, 369 
\bibitem[Machida et al.(2008)]{2008ApJ...676.1088M} Machida, M.~N., Inutsuka, S.-i., \& Matsumoto, T.\ 2008, \apj, 676, 1088-1108 
\bibitem[Machida et al.(2011)]{2011PASJ...63..555M} Machida, M.~N., Inutsuka, S.-I., \& Matsumoto, T.\ 2011, \pasj, 63, 555 
\bibitem[Machida \& Hosokawa(2013)]{2013MNRAS.431.1719M} Machida, M.~N., \& Hosokawa, T.\ 2013, \mnras, 431, 1719 
\bibitem[Matthews \& Wilson(2000)]{2000ApJ...531..868M} Matthews, B.~C., \& Wilson, C.~D.\ 2000, \apj, 531, 868 
\bibitem[Matthews et al.(2001)]{2001ApJ...562..400M} Matthews, B.~C., Wilson, C.~D., \& Fiege, J.~D.\ 2001, \apj, 562, 400 
\bibitem[Matthews et al.(2005)]{2005ApJ...626..959M} Matthews, B.~C., Lai, S.-P., Crutcher, R.~M., \& Wilson, C.~D.\ 2005, \apj, 626, 959 
\bibitem[Maury et al.(2018)]{2018MNRAS.tmp..552M} Maury, A.~J., Girart, J.~M., Zhang, Q., et al.\ 2018, \mnras,  
\bibitem[McMullin et al.(2007)]{2007ASPC..376..127M} McMullin, J.~P., Waters, B., Schiebel, D., Young, W., \& Golap, K.\ 2007, Astronomical Data Analysis Software and Systems XVI, 376, 127 
\bibitem[Mellon \& Li(2008)]{2008ApJ...681.1356M} Mellon, R.~R., \& Li, Z.-Y.\ 2008, \apj, 681, 1356-1376 
\bibitem[Menten et al.(2007)]{2007A&A...474..515M} Menten, K.~M., Reid, M.~J., Forbrich, J., \& Brunthaler, A.\ 2007, \aap, 474, 515 
\bibitem[Mouschovias \& Spitzer(1976)]{1976ApJ...210..326M} Mouschovias, T.~C., \& Spitzer, L., Jr.\ 1976, \apj, 210, 326 
\bibitem[Nakano(1988)]{1988PASJ...40..593N} Nakano, T.\ 1988, \pasj, 40, 593 
\bibitem[Nakazato et al.(2003)]{2003ApJ...583..322N} Nakazato, T., Nakamoto, T., \& Umemura, M.\ 2003, \apj, 583, 322 
\bibitem[Ohashi et al.(2018)]{2018ApJ...864...81O} Ohashi, S., Kataoka, A., Nagai, H., et al.\ 2018, \apj, 864, 81 
\bibitem[Poidevin et al.(2010)]{2010ApJ...716..893P} Poidevin, F., Bastien, P., \& Matthews, B.~C.\ 2010, \apj, 716, 893 
\bibitem[Rao et al.(1998)]{1998ApJ...502L..75R} Rao, R., Crutcher, R.~M., Plambeck, R.~L., \& Wright, M.~C.~H.\ 1998, \apjl, 502, L75 
\bibitem[Rao et al.(2009)]{2009ApJ...707..921R} Rao, R., Girart, J.~M., Marrone, D.~P., Lai, S.-P., \& Schnee, S.\ 2009, \apj, 707, 921 
\bibitem[Rao et al.(2014)]{2014ApJ...780L...6R} Rao, R., Girart, J.~M., Lai, S.-P., \& Marrone, D.~P.\ 2014, \apjl, 780, L6 
\bibitem[Reynolds(1986)]{1986ApJ...304..713R} Reynolds, S.~P.\ 1986, \apj, 304, 713 
\bibitem[Sadavoy et al.(2016)]{2016A&A...588A..30S} Sadavoy, S.~I., Stutz, A.~M., Schnee, S., et al.\ 2016, \aap, 588, A30 
\bibitem[Sadavoy et al.(2018)]{2018ApJ...859..165S} Sadavoy, S.~I., Myers, P.~C., Stephens, I.~W., et al.\ 2018, \apj, 859, 165 
\bibitem[Shu et al.(1987)]{1987ARA&A..25...23S} Shu, F.~H., Adams, F.~C., \& Lizano, S.\ 1987, \araa, 25, 23 
\bibitem[Spitzer \& Tukey(1949)]{1949Sci...109..461S} Spitzer, L., \& Tukey, J.~W.\ 1949, Science, 109, 461 
\bibitem[Stephens et al.(2017)]{2017ApJ...851...55S} Stephens, I.~W., Yang, H., Li, Z.-Y., et al.\ 2017, \apj, 851, 55 
\bibitem[Tang et al.(2013)]{2013ApJ...763..135T} Tang, Y.-W., Ho, P.~T.~P., Koch, P.~M., Guilloteau, S., \& Dutrey, A.\ 2013, \apj, 763, 135 
\bibitem[Takahashi et al.(2009)]{2009ApJ...704.1459T} Takahashi, S., Ho, P.~T.~P., Tang, Y.-W., Kawabe, R., \& Saito, M.\ 2009, \apj, 704, 1459 
\bibitem[Takahashi \& Ho(2012)]{2012ApJ...745L..10T} Takahashi, S., \& Ho, P.~T.~P.\ 2012, \apjl, 745, L10 
\bibitem[Takahashi et al.(2012)]{2012ApJ...752...10T} Takahashi, S., Saigo, K., Ho, P.~T.~P., \& Tomida, K.\ 2012, \apj, 752, 10 
\bibitem[Takahashi et al.(2013)]{2013ApJ...763...57T} Takahashi, S., Ho, P.~T.~P., Teixeira, P.~S., Zapata, L.~A., \& Su, Y.-N.\ 2013, \apj, 763, 57 
\bibitem[Tazaki et al.(2017)]{2017ApJ...839...56T} Tazaki, R., Lazarian, A., \& Nomura, H.\ 2017, \apj, 839, 56 
\bibitem[Tomida et al.(2010)]{2010ApJ...714L..58T} Tomida, K., Tomisaka, K., Matsumoto, T., et al.\ 2010, \apjl, 714, L58 
\bibitem[Tomida et al.(2017)]{2017ApJ...835L..11T} Tomida, K., Machida, M.~N., Hosokawa, T., Sakurai, Y., \& Lin, C.~H.\ 2017, \apjl, 835, L11 
\bibitem[Tomisaka(2002)]{2002ApJ...575..306T} Tomisaka, K.\ 2002, \apj, 575, 306 
\bibitem[Tomisaka(2011)]{2011PASJ...63..147T} Tomisaka, K.\ 2011, \pasj, 63, 147
\bibitem[Wilner \& Welch(1994)]{1994ApJ...427..898W} Wilner, D.~J., \& Welch, W.~J.\ 1994, \apj, 427, 898 
\bibitem[Yang et al.(2016)]{2016MNRAS.460.4109Y} Yang, H., Li, Z.-Y., Looney, L.~W., et al.\ 2016, \mnras, 460, 4109 
\bibitem[Yang et al.(2017)]{2017MNRAS.472..373Y} Yang, H., Li, Z.-Y., Looney, L.~W., Girart, J.~M., \& Stephens, I.~W.\ 2017, \mnras, 472, 373 
\bibitem[Zhang et al.(2014)]{2014ApJ...792..116Z} Zhang, Q., Qiu, K., Girart, J.~M., et al.\ 2014, \apj, 792, 116 
\end{thebibliography}
\end{document}